\documentclass[11pt,a4paper]{article}

\usepackage[T1]{fontenc}
\usepackage[utf8]{inputenc}
\usepackage{lmodern}
\usepackage[a4paper,margin=1in]{geometry}
\usepackage{amsmath,amssymb,amsthm,mathtools}
\usepackage{bm}
\usepackage{mathrsfs}
\usepackage{enumitem}
\usepackage{array,booktabs,longtable}
\usepackage{microtype}

\makeatletter
\@ifundefined{ifGin@setpagesize}{%
  \newif\ifGin@setpagesize
  \Gin@setpagesizetrue
}{}
\makeatother
\usepackage{graphicx}
\usepackage{xcolor}
\usepackage{tikz}
\usetikzlibrary{arrows.meta}
\usepackage[colorlinks=true,linkcolor=blue,urlcolor=blue,citecolor=magenta]{hyperref}

\numberwithin{equation}{section}
\allowdisplaybreaks

\newcommand{\ii}{\mathrm{i}}
\newcommand{\dd}{\mathrm{d}}
\newcommand{\RR}{\mathbb{R}}
\newcommand{\RP}{\mathbb{RP}}
\newcommand{\CN}{Cayley--Niederer}

\begin{document}

\begin{center}
{\Large\bfseries The Free Particle--Oscillator--Inverted Oscillator Triangle:\\[2pt]
Conformal Bridges, Metaplectic Rotations\\ [3pt] and \(\mathfrak{osp}(1|2)\) Structure}

\vspace{6pt}
{\bfseries Andrey Alcala and Mikhail S. Plyushchay}\\[8pt]
{\small\itshape Departamento de F\'{\i}sica, Universidad de Santiago de Chile,\\
Av. Victor Jara 3493, Santiago, Chile}\\[4pt]
{\small\slshape E-mails: \textcolor{blue}{andrey.alcala@usach.cl},
\textcolor{blue}{mikhail.plyushchay@usach.cl}}
\end{center}

\begin{abstract}
We study the free particle (FP), harmonic oscillator (HO) and
inverted harmonic oscillator (IHO) as parabolic, elliptic and
hyperbolic realizations of one conformal/metaplectic structure.
Within this framework, we develop a common free-particle
construction of HO bound and coherent states and IHO resonance
and scattering data.
The construction extends the FP--HO conformal bridge to both hyperbolic
orientations using real metaplectic rotations through $\pm\pi/4$.
These rotations map $2\Omega D$ to $\mp H_{\rm IHO}$ and carry
the FP threshold Jordan module into the two Gamow families. Applied to
Fourier characters, they give light-cone generating states whose Taylor
and Mellin decompositions recover resonance and real-energy connection
data, respectively. The familiar IHO scattering amplitudes and their
Gamma-function poles thereby enter the same construction. We identify
the crossed roles of the elliptic and hyperbolic generators in the two
stationary bridges and develop the hyperbolic Cayley--Niederer map as
their time-dependent counterpart. The $\mathfrak{osp}(1|2)$ extension
keeps track of the two parity sectors and the linear generating operators.
A phase-space analysis reconciles exact Liouville transport with nonzero
stationary transmission, distinguishing asymptotic quantum currents from
classical trajectories across separatrices. 
We also briefly discuss applications of the common quadratic structure to
barrier scattering, time-dependent oscillators and quantum fields
in gravitational backgrounds.
\end{abstract}

\bigskip

\section{Introduction}
\label{sec:introduction}

The free particle (FP), harmonic oscillator (HO) and inverted harmonic
oscillator (IHO) are the elementary parabolic, elliptic and hyperbolic
quadratic systems. The HO supplies the stable normal form and the Fock,
coherent-state and squeezed-state constructions used throughout quantum
theory~\cite{Glauber1963,Perelomov,WallsMilburn1994}. Broader contexts
include entanglement entropy~\cite{Srednicki}, matrix
theory~\cite{BFSS} and bounds on quantum chaos~\cite{MalShenStan}.
The IHO is the elementary local model of hyperbolic motion near a
nondegenerate saddle in phase space. The quantum transmission problem
for an inverted parabolic barrier goes back to Kemble's pioneering
work~\cite{Kemble1935}; see also~\cite{LandauLifshitzQM1977}.
Applications of the IHO include barrier and
quantum Hall scattering, Schwinger production, Rindler/Unruh and Hawking
settings, black-hole barrier approximations, matrix models, and
Berry--Keating spectral problems~\cite{Barton1990,BalazsVoros1990,SubramanyanHegdeVishveshwaraBradlyn2021,Schwinger1951,Fulling1973,Davies1975,Unruh1976,Hawking1975,Crispino2008,DouglasKlebanovKutasovMaldacenaMartinecSeiberg2005,BetziosGaddamPapadoulaki2016,Bhattacharyya2021,QuJiangLiu2022,BerryKeating1999,BerryKeating99+,SundaramBurgessODell2024}.
The FP also has a constructive role: it seeds Darboux transformations
to reflectionless systems and the associated KdV soliton
hierarchies~\cite{KayMoses,GGKM1967,Lax1968,SolNovikov,FadTakh,MatveevSalle1991,AblowitzClarkson1991,SolBelokolos}.

Their common quadratic algebra is
$\mathfrak{sl}(2,\mathbb R)\simeq\mathfrak{sp}(2,\mathbb R)$,
familiar from conformal mechanics, projective differential equations,
Virasoro/KdV geometry and holographic
correspondences~\cite{DeAlfaroFubiniFurlan1976,Schwarz1869,KhesinMisiolek2003,Maldacena1998,GubserKlebanovPolyakov1998,Witten1998}.
The Schr\"odinger and Newton--Hooke symmetries, and the maps between
free and oscillator evolution, are established
structures~\cite{LevyLeblond1963,Niederer1972,Hagen1972,Jackiw1972,Niederer1973,BacryLevyLeblond1968,DeromeDubois1972,Dubois1973,DuvalHorvathy2009}.
Their quantum implementation uses the metaplectic representation with
even and odd sectors $D^+_{1/4}\oplus D^+_{3/4}$; adjoining the linear
canonical generators gives the $\mathfrak{osp}(1|2)$
extension~\cite{Perelomov,Folland1989,Bargmann1961,MSPTorsion,MSPsl2R,InzMSPosp}.
These ingredients underlie the present construction and are reviewed
where needed to fix conventions.

Our starting point is the conformal-bridge
program~\cite{CBT1,CBTMono,InzMSPString,CBT5,CBT7} and its projective-time
formulation in Ref.~\cite{AlcalaPlyushchay2602.06378}. That work developed
the FP--HO stationary and time-dependent correspondences, identified the
Cayley/Bargmann interpretation, and included the hyperbolic time map in
the general Schwarzian construction. Here we develop the IHO state
construction and its relation to scattering, completing the triangle at
the level of threshold, generating-function and connection data.
Related conformal bridges and hidden symmetries occur in black-hole
mechanics~\cite{AchourLivine2021,AchourLivine2022,JaramilloLenziSopuerta2024}.

The IHO--dilation rotation, its Mellin eigenfunctions and the standard
transmission and reflection coefficients are also established
results~\cite{Barton1990,BalazsVoros1990,SubramanyanHegdeVishveshwaraBradlyn2021}.
Recent horizon and quadratic-gravity studies use this structure with
additional sector and quantization
prescriptions~\cite{GaztanagaKumarMarto2026,KumarMarto2026,KumarMarto2024Hawking}.
We compare these prescriptions with the ordinary whole-line IHO,
as well as with water-wave phase-space horizons and time-dependent
oscillator inversion~\cite{Rozenman2024,RajeevChakrabortyPadmanabhan2018}.
These comparisons clarify the common quadratic structure and
the additional physical input required in each application.

\paragraph{Main results.}
Our contribution is the explicit construction of the FP--HO--IHO state
dictionary from the same free-particle conformal module. Its stationary
hyperbolic part keeps both real rotations,
\[
 R_{-\pi/4}(2\Omega D)R_{-\pi/4}^{-1}=H_{\rm IHO},\qquad
 R_{+\pi/4}(2\Omega D)R_{+\pi/4}^{-1}=-H_{\rm IHO},
\]
where $D=(qp+pq)/4$ and
$R_\theta=\exp[-i\theta H_{\rm rot}/(\hbar\Omega)]$, with
$H_{\rm rot}=(p^2+\Omega^2q^2)/2$.
Applied to the FP threshold monomials, these give
$\Gamma_n^{(\pm)}=R_{\pm\pi/4}q^n$ with
$E_n^{\pm}=\pm i\hbar\Omega(n+\tfrac12)$,
as generalized resonance vectors with the corresponding boundary
prescriptions; the FP--HO bridge instead gives bound states.
The conjugated operator is the dilation generator, not the free
Hamiltonian, so this construction does not identify their self-adjoint
spectra. The two bridges have crossed generators: the FP--HO map uses
a complexified hyperbolic flow, whereas the FP--IHO map uses a real
elliptic rotation (Sections~\ref{sec:conformal-bridges-revised-improved}--\ref{sec:light-cone-gamow-revised}).

We then carry the FP Fourier character into a light-cone generating
state. Its Taylor expansion organizes the Gamow tower, while its Mellin
decomposition yields the real-energy connection coefficients
\eqref{eq:sec6-v2-tr-amplitudes}; their Gamma-function continuation
recovers the same resonance poles (Section~\ref{sec:scattering-Fourier-Mellin-revised-v2}).
Thus the known IHO amplitudes are derived within the same construction
as the FP--HO coherent-state map. The linear generators connect the two
parity sectors and make the $\mathfrak{osp}(1|2)$ organization explicit (Section~\ref{sec:algebraic-backbone}).
We complete this stationary dictionary with its time-dependent
hyperbolic Cayley--Niederer realization,
$t=\Omega^{-1}\tanh(\Omega\tau)$,
including the wave-function map and propagator
(Section~\ref{sec:hyperbolic-CN-TDSE}).

The interpretation of the connection matrix requires a further
distinction. Incoming and outgoing light-cone bases belong to conjugate
polarizations; their connection coefficients give asymptotic
coordinate-space fluxes, not a classical trajectory crossing a
phase-space separatrix. Sections~\ref{sec:phase-space-wigner}--\ref{sec:physical-applications-final-style-v2}
and Appendix~\ref{app:weber-iho} explain this distinction and its physical
scope. The remaining appendices collect the finite-order conventions,
the Berry--Keating/inverse-square relation, and the detailed
Perelomov-state construction. 
Section~ \ref{sec:discussion-outlook}  presents the concluding discussion and outlook.

\section{Algebraic and representation-theoretic backbone}
\label{sec:algebraic-backbone}

The free particle (FP), harmonic oscillator (HO), and inverted harmonic oscillator (IHO) are the parabolic, elliptic, and hyperbolic realizations of one quadratic conformal algebra,
\(\mathfrak{sl}(2,\mathbb R)\simeq\mathfrak{sp}(2,\mathbb R)\simeq\mathfrak{su}(1,1)\),
and, after quantization, of the same metaplectic representation. Their distinct self-adjoint spectra, recalled below, preclude a unitary identification of the three Hamiltonians.

\subsection{Quadratic conformal algebra}
\label{subsec:quadratic-conformal-algebra}

We use one canonical pair, \([q,p]=i\hbar\), and the quadratic generators
\begin{equation}
H_0=\frac12 p^2,\qquad K=\frac12 q^2,\qquad D=\frac14(qp+pq).
\label{eq:H0-K-D-sec2-rev}
\end{equation}
With this normalization,
\begin{equation}
[D,H_0]=i\hbar H_0,\qquad [D,K]=-i\hbar K,
\qquad [H_0,K]=-2i\hbar D.
\label{eq:sl2-H0-K-D-sec2-rev}
\end{equation}
This convention is adapted to the Bargmann weights.  Indeed, in the coordinate representation
\(p=-i\hbar d/dq\) and
\(D=-(i\hbar/2)(q\,d/dq+1/2)\), so that
\begin{equation}
\frac{i}{\hbar}Dq^n=\left(\frac n2+\frac14\right)q^n .
\label{eq:iD-qn-sec2-rev}
\end{equation}
Consequently the monomials \(q^{2m}\) and \(q^{2m+1}\) have weights \(m+1/4\) and \(m+3/4\), respectively.
 When the twice-normalized dilation generator is needed, it will be written explicitly as \(2D=(qp+pq)/2\); this avoids introducing a second notation and keeps the factors of two visible in the FP--IHO bridge.

The oscillator and inverted-oscillator Hamiltonians are the elliptic and hyperbolic combinations
\begin{equation}
H_{\rm HO}=H_+=H_0+\omega^2K=\frac12p^2+\frac12\omega^2q^2,
\qquad
H_{\rm IHO}=H_-=H_0-\Omega^2K=\frac12p^2-\frac12\Omega^2q^2 .
\label{eq:Hplus-Hminus-dim-sec2-rev}
\end{equation}
In units $\omega=\Omega=1$, this reduces to \(H_+=(p^2+q^2)/2\) and \(H_-=(p^2-q^2)/2\), with
\begin{equation}
[D,H_+]=i\hbar H_-,\qquad [D,H_-]=i\hbar H_+,
\qquad [H_+,H_-]=4i\hbar D.
\label{eq:D-Hplus-Hminus-sec2-rev}
\end{equation}
Thus \(H_0\) and \(K\) are parabolic generators, \(H_+\) is elliptic, and \(H_-\) and \(D\) are hyperbolic.  The physical triangle is formed by \(H_0\), \(H_+\), and \(H_-\); the generator \(K\) completes the conformal triple but is not treated here as a fourth Hamiltonian on the same footing.\footnote{The generator \(K=\frac12 q^2\) is unitarily equivalent to \(H_0=\frac12 p^2\) by Fourier transformation, and hence is parabolic as an \(\mathfrak{sl}(2,\mathbb R)\) generator.  Its Hamiltonian flow, however, is qualitatively different: \(\dot q=0\), \(\dot p=-q\).  In this sense it is naturally related to Carrollian or ultralocal dynamics, where spatial motion is frozen.  Such \(K\)-type or Carrollian structures appear in Carrollian and ultralocal dynamics, including Carroll particles and Carroll quantum Hall systems~\cite{Bergshoeff2014,Marsot2021,FOFPP,Ruzziconi}.}

\subsection{Different spectra, same metaplectic representation}
\label{subsec:different-spectra-same-rep}

In their standard self-adjoint realizations,
\begin{equation}
\sigma(H_{\rm FP})=[0,\infty),\qquad
\sigma(H_{\rm HO})=\left\{\hbar\omega\left(n+\frac12\right):n=0,1,2,\ldots\right\},
\qquad
\sigma(H_{\rm IHO})=\mathbb R .
\label{eq:spectra-sec2-rev}
\end{equation}
Hence the bridges discussed below cannot be ordinary unitary equivalences of self-adjoint Hamiltonians.  They compare different spectral realizations of the same metaplectic representation.  Diagonalizing \(H_0\) gives the parabolic/free-particle basis; diagonalizing \(H_+\) gives the elliptic/oscillator basis; diagonalizing \(H_-\) gives the hyperbolic/IHO scattering basis; and diagonalizing \(D\) gives the dilation, or Mellin, basis.  The HO and IHO structures are therefore not hidden in the numerical spectrum of \(H_0\) alone, but in the full conformal module carried by the free particle.

\subsection{Oscillator realization and the two Bargmann sectors}
\label{subsec:oscillator-Bargmann-sectors-rev}

The oscillator realization displays the representation content most transparently.  With
\begin{equation}
 a^- = \sqrt{\frac{\omega}{2\hbar}}q+\frac{i}{\sqrt{2\hbar\omega}}p,
\qquad
 a^+ = \sqrt{\frac{\omega}{2\hbar}}q-\frac{i}{\sqrt{2\hbar\omega}}p,
\label{eq:a-pm-sec2-rev}
\end{equation}
one has \([a^-,a^+]=1\) and \(H_{\rm HO}=\hbar\omega(a^+a^-+1/2)\).  The quadratic combinations
\begin{equation}
J_0=\frac12\left(a^+a^-+\frac12\right),\qquad
J_+=\frac12(a^+)^2,
\qquad
J_- = \frac12(a^-)^2
\label{eq:J-sec2-rev}
\end{equation}
satisfy \([J_0,J_\pm]=\pm J_\pm\), \([J_-,J_+]=2J_0\), and \(H_{\rm HO}=2\hbar\omega J_0\).  The even oscillator states \(|0\rangle,|2\rangle,\ldots\) carry \(D^+_{1/4}\), whereas the odd states \(|1\rangle,|3\rangle,\ldots\) 
carry \(D^+_{3/4}\)~\cite{Perelomov,Folland1989,Bargmann1961}.  Indeed,
\begin{equation}
J_0|2m\rangle=\left(m+\frac14\right)|2m\rangle,
\qquad
J_0|2m+1\rangle=\left(m+\frac34\right)|2m+1\rangle.
\label{eq:J0-weights-sec2-rev}
\end{equation}
The common metaplectic representation is therefore
\begin{equation}
D^+_{1/4}\oplus D^+_{3/4}.
\label{eq:metaplectic-decomposition-sec2-rev}
\end{equation}
For the Casimir convention
\begin{equation}
\mathcal C_{\mathfrak{su}(1,1)}=J_0^2-\frac12(J_+J_-+J_-J_+),
\label{eq:su11-Casimir-sec2-rev}
\end{equation}
both sectors give \(\mathcal C_{\mathfrak{su}(1,1)}=k(k-1)=-3/16\), for \(k=1/4\) and \(k=3/4\).  The equality of the Casimir values is essential: the two sectors are inequivalent as \(\mathfrak{su}(1,1)\) lowest-weight modules, but they are the two parity sectors of the same one-dimensional metaplectic representation.

\subsection{The free-particle threshold module}
\label{subsec:FP-threshold-module-rev}

The same weights appear on the free-particle side.  In coordinate representation \(H_0=-(\hbar^2/2)d^2/dq^2\), and the monomials \(1,q,q^2,q^3,\ldots\) are generalized, non-normalizable states in the rigged-Hilbert-space sense~\cite{RiggedHil}.  They form two zero-energy Jordan chains because
\begin{equation}
H_0q^n=-\frac{\hbar^2}{2}n(n-1)q^{n-2},
\label{eq:H0-qn-sec2-rev}
\end{equation}
so that the even chain starts from \(1\) and the odd chain from \(q\).  
At the same time, \(Kq^n=\frac12 q^{n+2}\), so that \(K\) raises the degree by two, while Eq.~\eqref{eq:iD-qn-sec2-rev} gives
\begin{equation}
\frac{i}{\hbar}Dq^{2m}=\left(m+\frac14\right)q^{2m},
\qquad
\frac{i}{\hbar}Dq^{2m+1}=\left(m+\frac34\right)q^{2m+1}.
\label{eq:iD-even-odd-sec2-rev}
\end{equation}
Thus the free-particle threshold sector already carries the two lowest-weight patterns which become the even and odd oscillator towers after the FP--HO bridge, and the two parity subsectors within each Gamow family after the FP--IHO bridge.

The momentum operator \(p=-i\hbar d/dq\) is a true integral of \(H_0\), \([H_0,p]=0\).\footnote{Unlike \(p\), the coordinate \(q\) is not itself a time-independent integral of the free-particle Hamiltonian. It is promoted to the Galilean boost integral \(X(t)=q-tp\). Similarly, the \(t=0\) generators \(D\) and \(K\) are promoted to the dynamical conformal integrals \(D(t)=D-tH_0\) and \(K(t)=K-2tD+t^2H_0\), for the normalization \(D=(qp+pq)/4\).}
Hence the plane waves \(\psi_k(q)=\exp(ikq)\) satisfy \(p\psi_k=\hbar k\psi_k\) and \(H_0\psi_k=(\hbar^2k^2/2)\psi_k\).  Their expansion
\begin{equation}
\exp(ikq)=\sum_{n=0}^{\infty}\frac{(ik)^n}{n!}q^n
\label{eq:plane-wave-Taylor-sec2-rev}
\end{equation}
shows that the Jordan states are the Taylor jets of the free-particle continuum at \(k=0\), or equivalently at the scale-fixed threshold \(E=0\).  The constant state is also the kernel of the first-order integral, \(p\,1=0\).  
This first-order structure is one of the reasons why the superconformal extension is natural here.

\subsection{Linear generators and the \texorpdfstring{$\mathfrak{osp}(1|2)$}{osp(1|2)} extension}
\label{subsec:osp-extension-sec2-rev}

The quadratic generators form \(\mathfrak{sl}(2,\mathbb R)\).  Including the linear canonical generators gives the natural superconformal extension \(\mathfrak{osp}(1|2)\).  This is not only a formal enlargement: the linear generators connect the two metaplectic sectors and their eigenstates generate the coherent-state and scattering-state structures used later.

To avoid ambiguity, Poisson brackets will be denoted by \(\{\ ,\ \}\) only in explicitly classical formulas, while operator anticommutators are written as \([A,B]_+=AB+BA\).  In the free-particle realization the odd generators are \(q,p\), and
\begin{equation}
[p,p]_+=4H_0,
\qquad
[q,q]_+=4K,
\qquad
[q,p]_+=4D.
\label{eq:osp-FP-sec2-rev}
\end{equation}
In the oscillator realization one may use \(a^+,a^-\), or, more symmetrically, \(F_+=a^+/\sqrt2\), \(F_-=a^-/\sqrt2\).  Together with \(J_0,J_\pm\), these obey
\begin{equation}
[J_0,F_\pm]=\pm\frac12F_\pm,
\qquad
[J_+,F_-]=-F_+,
\qquad
[J_-,F_+]=F_-,
\label{eq:osp-even-odd-sec2-rev}
\end{equation}
with \([J_+,F_+]=[J_-,F_-]=0\), and
\begin{equation}
[F_+,F_+]_+=2J_+,
\qquad
[F_-,F_-]_+=2J_-,
\qquad
[F_+,F_-]_+=2J_0.
\label{eq:osp-odd-odd-sec2-rev}
\end{equation}
The corresponding \(\mathfrak{osp}(1|2)\) Casimir may be chosen as
\begin{equation}
\mathcal C_{\mathfrak{osp}(1|2)}
=J_0^2-\frac12(J_+J_-+J_-J_+)-\frac14(F_+F_- - F_-F_+).
\label{eq:osp-Casimir-sec2-rev}
\end{equation}
For the oscillator/metaplectic module considered here, \(F_+F_- -F_-F_+=-1/2\), and therefore
\begin{equation}
\mathcal C_{\mathfrak{osp}(1|2)}=-\frac{1}{16}.
\label{eq:osp-Casimir-value-sec2-rev}
\end{equation}

The \(\mathbb Z_2\)-grading is supplied by parity, \(\Gamma\psi(q)=\psi(-q)\), or equivalently 
\(\Gamma=(-1)^N=\cos(\pi N)\) in oscillator notation.  The quadratic generators commute with \(\Gamma\), while the linear ones anticommute with it.  Hence \(D^+_{1/4}\oplus D^+_{3/4}\) is reducible as an \(\mathfrak{sl}(2,\mathbb R)\) representation, but becomes the natural irreducible metaplectic/superconformal module once the odd generators are included.

This has direct physical meaning.  The operator \(p\) is both a true integral of the free particle and an odd generator of the superconformal extension.  Its eigenstates are the plane waves.  Under the FP--HO bridge these states become oscillator coherent states; under the FP--IHO bridge they are reorganized into light-cone coherent states and, after Mellin decomposition, into IHO scattering states.  Thus \(\mathfrak{osp}(1|2)\) keeps track of the first-order operators whose eigenstates generate the coherent and scattering data.

\subsection{Hyperbolic realization and real metaplectic rotations}
\label{subsec:hyperbolic-quarter-sec2-rev}

For the IHO in unit-frequency notation define
\begin{equation}
u_+=\frac{p+q}{\sqrt2},
\qquad
u_- = \frac{p-q}{\sqrt2},
\qquad
[u_+,u_-]=i\hbar.
\label{eq:u-pm-sec2-rev}
\end{equation}
Then
\begin{equation}
H_- = \frac12(p^2-q^2)=\frac12(u_+u_-+u_-u_+),
\qquad
H_{\rm IHO}=\frac{\Omega}{2}(u_+u_-+u_-u_+).
\label{eq:Hminus-u-sec2-rev}
\end{equation}
In the frequency-restored formulas, \(u_\pm\) use the scaled canonical pair
\(Q=\sqrt{\Omega}\,q\), \(P=p/\sqrt{\Omega}\) of
\eqref{eq:Hrot-HIHO-D-sec4-v2}.
The Heisenberg equations are \(\dot{u}_+=\Omega u_+\), \(\dot{u}_-=-\Omega u_-\).  Thus the IHO flow is the hyperbolic, boost-like flow in the two light-cone directions.  In the \(u_+\)-polarization, \(u_+=u\), \(u_-=-i\hbar d/du\), so
\begin{equation}
H_{\rm IHO}=-i\hbar\Omega\left(u\frac{d}{du}+\frac12\right).
\label{eq:HIHO-dilation-form-sec2-rev}
\end{equation}
The real-energy eigenfunctions are therefore Mellin-type distributions, \(u^{-1/2+iE/(\hbar\Omega)}\), with half-line support and boundary-value prescriptions fixed later in the scattering analysis.

The same variables are the odd generators of the hyperbolic realization.  They satisfy, for example, 
\([u_+,u_+]_+=2u_+^2\), \([u_-,u_-]_+=2u_-^2\), and \([u_+,u_-]_+=2H_-\).  They also shift IHO energy by imaginary steps:
\begin{equation}
[H_{\rm IHO},u_+]=-i\hbar\Omega u_+,
\qquad
[H_{\rm IHO}, u_-]=+i\hbar\Omega u_-.
\label{eq:u-ladder-sec2-rev}
\end{equation}
This is the hyperbolic analogue of the oscillator ladder structure.

The relation between the dilation generator and the IHO Hamiltonian is implemented by a real metaplectic rotation generated by \(H_+\).  With
\(\mathcal R_\theta=\exp(-i\theta H_+/\hbar)\), one has
\(\mathcal R_\theta q\mathcal R_\theta^{-1}=q\cos\theta-p\sin\theta\) and
\(\mathcal R_\theta p\mathcal R_\theta^{-1}=p\cos\theta+q\sin\theta\).  At the two angles \(\theta=\pm\pi/4\),
\begin{equation}
\mathcal R_{-\pi/4}(2D)\mathcal R_{-\pi/4}^{-1}=H_-,
\qquad
\mathcal R_{+\pi/4}(2D)\mathcal R_{+\pi/4}^{-1}=-H_-.
\label{eq:quarter-rotations-sec2-rev}
\end{equation}
With frequency restored, this becomes \(2\Omega D\mapsto \pm H_{\rm IHO}\).  Both signs must be kept: they correspond to opposite hyperbolic orientations, to the two light-cone polarizations, and to the two Gamow families \(E_n^\pm=\pm i\hbar\Omega(n+1/2)\).

The same passage can also be viewed formally at the level of first-order generators: under \(\omega\mapsto\pm i\Omega\), the oscillator combinations \(p\mp i\omega q\) become real combinations of \(p\) and \(q\), i.e. light-cone variables up to normalization and sign.  This is only an algebraic guide, not a Hilbert-space equivalence, but it explains why the elliptic odd generators \(a^\pm\) and the hyperbolic odd generators \(u_\pm\) belong to the same \(\mathfrak{osp}(1|2)\) picture.

\section{The free particle as the parabolic origin}
\label{sec:free-particle-parabolic-origin}

Two pieces of free-particle data are especially important: the threshold Jordan module at \(E=0\), and the momentum eigenstates, i.e. the plane waves.  The first one is the discrete seed for the oscillator bound states and the IHO Gamow states.  The second one is the generating object behind oscillator coherent states and the real-energy scattering data of the IHO.

\subsection{The scale-fixed threshold}
\label{subsec:scale-fixed-threshold}

In coordinate representation,
\begin{equation}
H_0=-\frac{\hbar^2}{2}\frac{d^2}{dq^2},
\qquad
H_0\psi=E\psi,
\qquad
E=\frac{\hbar^2k^2}{2}\geq 0 .
\label{eq:free-SSE-spectrum-sec3-rev}
\end{equation}
The point \(E=0\) is the threshold of the continuous spectrum.  It is also the unique fixed point of the dilation action.  Indeed, \([D,H_0]=i\hbar H_0\) implies
\begin{equation}
\exp\left(-\frac{i\alpha D}{\hbar}\right)H_0
\exp\left(\frac{i\alpha D}{\hbar}\right)=e^\alpha H_0.
\label{eq:dilation-energy-scaling-sec3-rev}
\end{equation}
The active state transformation
\(\psi_E\mapsto e^{-i\alpha D/\hbar}\psi_E\) sends its energy under
\(H_0\) to \(e^{-\alpha}E\). Thus all positive energies are moved along
the continuum, whereas \(E=0\) remains fixed. This double role, as
spectral threshold and scale-fixed point, makes the zero-energy sector
the natural conformal seed of the bridge construction.

At the threshold, \(H_0\psi=0\) gives \(\psi''(q)=0\), hence
\begin{equation}
\ker H_0=\operatorname{span}\{1,q\}
\label{eq:ker-H0-sec3-rev}
\end{equation}
in the generalized sense.  The momentum operator \(p=-i\hbar d/dq\) is a true first-order integral, \([H_0,p]=0\), but its zero mode is only \(\ker p=\operatorname{span}\{1\}\).  Thus the constant state is distinguished as the common zero mode of \(p\) and \(H_0\), while \(q\) belongs to the zero-energy conformal sector but not to the kernel of \(p\).

\subsection{Jordan chains and Bargmann weights}
\label{subsec:jordan-chains-bargmann-weights}

Each monomial \(q^n\), \(n=0,1,2,\ldots\), is annihilated by a finite power of \(H_0\). By \eqref{eq:H0-qn-sec2-rev}, the even and odd monomials form the two chains
\begin{equation}
1\leftarrow q^2\leftarrow q^4\leftarrow\cdots,
\qquad
q\leftarrow q^3\leftarrow q^5\leftarrow\cdots,
\label{eq:even-odd-Jordan-chains-sec3-rev}
\end{equation}
where the arrows denote the lowering action of \(H_0\), up to numerical factors.  The detailed normalization of the chains is not essential here; it can be chosen so that \(H_0\chi_n=\chi_{n-2}\) for \(n\geq2\).

The dilation weights in \eqref{eq:iD-even-odd-sec2-rev} identify these chains with the two lowest-weight patterns \(D^+_{1/4}\) and \(D^+_{3/4}\), which appear as the even and odd Fock towers in the oscillator realization.

The full polynomial threshold module is stable under the conformal
action. Each generalized kernel \(\ker H_0^r\) is preserved by \(D\),
whereas \(K\) maps it into \(\ker H_0^{r+1}\). Thus these kernels form
a filtration of the conformally invariant module
\(\bigcup_{r\geq1}\ker H_0^r\), rather than separate conformally
invariant sectors.

The linear generators act on the same module as
\begin{equation}
q\,q^n=q^{n+1},
\qquad
p\,q^n=-i\hbar n q^{n-1}.
\label{eq:linear-action-monomials-sec3-rev}
\end{equation}
Together with the quadratic generators they give the free-particle realization of the \(\mathfrak{osp}(1|2)\) structure described in Section~\ref{sec:algebraic-backbone}: the linear operators are odd with respect to parity, while the quadratic generators are even.

\subsection{Plane waves and additive characters}
\label{subsec:plane-waves-additive-characters}

For \(E>0\), the free-particle generalized eigenstates may be chosen as simultaneous eigenstates of \(H_0\) and the true integral \(p\):
\begin{equation}
\psi_k(q)=e^{ikq},
\qquad
p\psi_k=\hbar k\psi_k,
\qquad
H_0\psi_k=\frac{\hbar^2k^2}{2}\psi_k .
\label{eq:plane-waves-sec3-rev}
\end{equation}
The two signs of \(k\) distinguish the two branches of the positive-energy continuum, while at \(k=0\) the two branches meet at the constant threshold state.

The plane waves are also additive characters of translations,
\begin{equation}
\psi_k(q+a)=e^{ika}\psi_k(q).
\label{eq:additive-character-sec3-rev}
\end{equation}
This simple character property is one of the reasons why the free-particle realization plays a privileged role.  Under the FP--IHO bridge, the additive Fourier characters will be reorganized into multiplicative Mellin characters of the IHO light-cone variable.  The scattering coefficients are then read as Fourier--Mellin connection data.

\subsection{Threshold Taylor jets and generating functions}
\label{subsec:threshold-Taylor-jets}

The Taylor expansion \eqref{eq:plane-wave-Taylor-sec2-rev} relates the Jordan module to the plane-wave sector through
\begin{equation}
q^n=i^{-n}\left.\frac{d^n}{dk^n}e^{ikq}\right|_{k=0}.
\label{eq:plane-wave-Taylor-sec3-rev}
\end{equation}
Hence the free-particle Jordan states are the threshold Taylor jets of the continuum eigenfunctions.  Equivalently, suitable polynomially renormalized \(k\to0\) combinations of plane waves produce the monomials; for instance
\begin{equation}
1=\lim_{k\to0}\cos(kq),
\quad
q=\lim_{k\to0}\frac{\sin(kq)}{k},\quad
q^2=2\lim_{k\to0}\frac{1-\cos(kq)}{k^2},
\quad
q^3=6\lim_{k\to0}\frac{kq-\sin(kq)}{k^3}.
\label{eq:first-threshold-limits-sec3-rev}
\end{equation}
This generating-function viewpoint is central. If \(\mathcal B_+\) denotes the FP--HO bridge, the corresponding formal generating-function identity is
\begin{equation}
\mathcal B_+e^{ikq}=\sum_{n=0}^{\infty}\frac{(ik)^n}{n!}\mathcal B_+q^n .
\label{eq:Bplus-generating-sec3-rev}
\end{equation}
The map from FP plane waves to HO coherent states is therefore the summed, generating-function form of the map from FP Jordan states to oscillator bound states.

For the FP--IHO bridge, Taylor and Mellin decompositions of the same light-cone generating state yield, respectively,
\begin{equation}
\begin{aligned}
\text{Taylor expansion} &\longrightarrow \text{Gamow/resonance data},\\
\text{Mellin expansion} &\longrightarrow \text{real-energy scattering data}.
\end{aligned}
\label{eq:Taylor-Mellin-distinction-sec3-rev}
\end{equation}
This distinction prevents confusion between the complex-energy Gamow states and the real-energy generalized eigenstates of the self-adjoint IHO Hamiltonian.

\subsection{The two chains prepared by the free particle}
\label{subsec:two-chains-prepared-by-FP}

The preceding discussion prepares the two chains used throughout the FP--HO--IHO triangle:
\begin{equation}
\text{FP threshold Jordan states}
\longrightarrow
\text{HO bound states}
\longrightarrow
\text{IHO Gamow states},
\label{eq:Jordan-HO-IHO-chain-sec3-rev}
\end{equation}
for the discrete/resonant structures, and
\begin{equation}
\text{FP plane waves}
\longrightarrow
\text{HO coherent states}
\longrightarrow
\text{IHO scattering states},
\label{eq:plane-wave-HO-IHO-chain-sec3-rev}
\end{equation}
for the generating and scattering structures. The Taylor identity \eqref{eq:plane-wave-Taylor-sec2-rev} links the two chains. The monomials and plane waves are generalized states in the rigged-Hilbert-space setting, rather than normalizable vectors in \(L^2(\mathbb R,dq)\); their bridge images supply the bound, coherent, Gamow, and scattering structures of the triangle.

We now apply the two conformal bridges to the FP threshold module and
Fourier characters.

\section{Conformal bridges: FP--HO similarity and FP--IHO metaplectic rotation}
\label{sec:conformal-bridges-revised-improved}

The preceding sections identified the free-particle threshold module and the free plane waves as the basic parabolic data.  We now recall the already established FP--HO conformal bridge and develop its hyperbolic FP--IHO counterpart.  The stationary bridges do not identify the self-adjoint Hamiltonians as spectra-preserving operators.  Rather, they act on the dilation generator inside the full conformal module:
\begin{equation}
2\omega iD\longrightarrow H_{\rm HO},
\qquad
2\Omega D\longrightarrow \pm H_{\rm IHO}.
\label{eq:stationary-bridges-schematic-sec4-v2}
\end{equation}
The first map is a nonunitary similarity transformation in the Schr\"odinger representation and is generated by a complexified hyperbolic flow.  The second is a real unitary metaplectic rotation through \(\pm\pi/4\).  This crossed relation is one of the useful dualities of the FP--HO--IHO triangle.

\subsection{The FP--HO bridge}
\label{subsec:FP-HO-bridge-v2}

In unit-frequency notation, with \(\hbar\) retained, the Cayley form of the FP--HO bridge is generated by the hyperbolic operator
\(H_-=(p^2-q^2)/2\).  Up to an overall scalar normalization,
\begin{equation}
U_C \propto \exp\left(\frac{\pi}{4\hbar}H_-\right)
       =\exp\left[-\frac{i}{\hbar}\left(i\frac{\pi}{4}\right)H_-\right],
\label{eq:Cayley-operator-Hminus-sec4-v2}
\end{equation}
so the bridge is the IHO evolution analytically continued to the imaginary time parameter
\(t=i\pi/4\).  Its similarity action implements the complex canonical transformation
\begin{equation}
U_CqU_C^{-1}=\sqrt{\hbar}\,a^+,
\qquad
U_CpU_C^{-1}=-i\sqrt{\hbar}\,a^-,
\qquad
U_C(2iD)U_C^{-1}=H_+,
\label{eq:UC-main-maps-sec4-v2}
\end{equation}
where \([a^-,a^+]=1\) and
\(H_+=\hbar(a^+a^-+1/2)=(p^2+q^2)/2\).
Restoring the oscillator frequency gives the first relation in
\eqref{eq:stationary-bridges-schematic-sec4-v2}.

The same complex-metaplectic element has a useful factorized coordinate representative.  With
\(H_0=p^2/2\), \(K=q^2/2\), and \(D=(qp+pq)/4\), one may write, up to an irrelevant scalar,
\begin{equation}
U_{C,\omega}
\propto
\exp\left(-\frac{\omega}{\hbar}K\right)
\exp\left(\frac{1}{2\hbar\omega}H_0\right)
\exp\left(\frac{i\ln2}{\hbar}D\right)
\propto
\exp\left(\frac{\pi}{4\hbar\omega}(H_0-\omega^2K)\right).
\label{eq:UC-Somega-relation-sec4-v2}
\end{equation}
The last expression makes explicit that the generator is the dimensionful IHO combination
\(H_0-\omega^2K\).  If the final dilation factor in \eqref{eq:UC-Somega-relation-sec4-v2} is suppressed, one obtains the simpler representative
\begin{equation}
S_\omega=
\exp\left(-\frac{\omega}{\hbar}K\right)
\exp\left(\frac{1}{2\hbar\omega}H_0\right),
\label{eq:Somega-definition-sec4-v2}
\end{equation}
which is the form most convenient for the elementary coordinate-space maps
\begin{equation}
S_\omega pS_\omega^{-1}=p-i\omega q=-i\sqrt{2\hbar\omega}\,a^- ,
\qquad
S_\omega qS_\omega^{-1}=\frac12\left(q-\frac{i}{\omega}p\right)
=\sqrt{\frac{\hbar}{2\omega}}\,a^+ .
\label{eq:Somega-linear-map-sec4-v2}
\end{equation}
The two operators \(U_{C,\omega}\) and \(S_\omega\) therefore represent the same Cayley bridge, but with different normalizations of the linear generators.  The scalar normalization is fixed only after choosing a Hilbert-space convention, for example the Schr\"odinger normalization of oscillator eigenstates or the Bargmann-Fock normalization.

The physical content is transparent from \eqref{eq:Somega-linear-map-sec4-v2}.  If \(\psi_k(q)=\exp(ikq)\), then \(p\psi_k=\hbar k\psi_k\), and \(\Psi_k=S_\omega\psi_k\) obeys
\begin{equation}
a^-\Psi_k=\alpha(k)\Psi_k,
\qquad
\alpha(k)=\frac{i\hbar k}{\sqrt{2\hbar\omega}}.
\label{eq:plane-wave-to-HO-coherent-sec4-v2}
\end{equation}
Thus free plane waves are mapped to ordinary HO Glauber coherent states.  Similarly, the free threshold monomials are mapped to the oscillator Fock tower,
\(q^n\mapsto \psi_n^{\rm HO}\), with
\(H_{\rm HO}\psi_n^{\rm HO}=\hbar\omega(n+1/2)\psi_n^{\rm HO}\).  These two maps are linked by the generating-function identity \eqref{eq:Bplus-generating-sec3-rev}.

The bridge is a nonunitary similarity in the Schr\"odinger
representation, with the domain qualifications described in
Appendix~\ref{app:order-eight}. It sends generalized threshold states
to normalizable oscillator states and maps \(2iD\) to the compact
Hermitian generator \(H_+\).  The same Cayley transformation becomes unitary when interpreted as a change of polarization from the Schr\"odinger representation to the Bargmann-Fock representation, where \(a^+\) acts by multiplication by a holomorphic variable and \(a^-\) by differentiation.  Thus the FP--HO bridge has two complementary faces: a nonunitary stationary similarity transformation and a unitary Bargmann change of polarization~\cite{Folland1989,Bargmann1961,AlcalaPlyushchay2602.06378}.

\subsection{The FP--IHO bridge}
\label{subsec:FP-IHO-bridge-v2}

The hyperbolic bridge is different.  It relates the Hermitian dilation generator \(D\) to the Hermitian IHO generator by a real metaplectic rotation.  In frequency-scaled canonical variables \(Q=\sqrt{\Omega}\,q\), \(P=p/\sqrt{\Omega}\),
\begin{equation}
H_{\rm rot}=\frac{\Omega}{2}(P^2+Q^2),
\qquad
H_{\rm IHO}=\frac{\Omega}{2}(P^2-Q^2),
\qquad
D=\frac14(QP+PQ).
\label{eq:Hrot-HIHO-D-sec4-v2}
\end{equation}
The unitary operator
\(R_\theta=\exp[-i\theta H_{\rm rot}/(\hbar\Omega)]\) rotates phase space by the angle \(\theta\).  At \(\theta=\pm\pi/4\) it produces the light-cone variables
\begin{equation}
u_+=\frac{P+Q}{\sqrt2},
\qquad
u_- =\frac{P-Q}{\sqrt2},
\qquad
[u_+,u_-]=i\hbar,
\qquad
H_{\rm IHO}=\frac{\Omega}{2}(u_+u_-+u_-u_+).
\label{eq:light-cone-IHO-sec4-v2}
\end{equation}
The same rotation gives
\begin{equation}
R_{-\pi/4}(2\Omega D)R_{-\pi/4}^{-1}=H_{\rm IHO},
\qquad
R_{+\pi/4}(2\Omega D)R_{+\pi/4}^{-1}=-H_{\rm IHO}.
\label{eq:D-to-HIHO-quarter-rotations-sec4-v2}
\end{equation}
No complexification is involved: the bridge is a unitary metaplectic transformation applied to generalized free-particle threshold states. The two target generators have opposite
Hamiltonian flows in a fixed time orientation. Their use as IHO and
dual-IHO sectors in the quadratic-gravity proposal of
Ref.~\cite{KumarMarto2026} involves further physical assumptions;
here both signs are retained within the stationary bridge.

The two signs in \eqref{eq:D-to-HIHO-quarter-rotations-sec4-v2} are essential.  Since
\(2\Omega Dq^n=-i\hbar\Omega(n+1/2)q^n\), the states
\(\Gamma_n^{(-)}=R_{-\pi/4}q^n\) and \(\Gamma_n^{(+)}=R_{+\pi/4}q^n\) satisfy
\begin{equation}
H_{\rm IHO}\Gamma_n^{(-)}=-i\hbar\Omega\left(n+\frac12\right)\Gamma_n^{(-)},
\qquad
H_{\rm IHO}\Gamma_n^{(+)}=+i\hbar\Omega\left(n+\frac12\right)\Gamma_n^{(+)}.
\label{eq:Gamow-families-sec4-v2}
\end{equation}
These are the two Gamow families. Their equally spaced imaginary energies give quantized decay or growth rates, with the interpretation fixed by the resonant or anti-resonant boundary condition. The image of a free plane wave is an eigenstate of a linear light-cone operator: a generalized hyperbolic coherent state, not a normalizable Glauber state. Its Taylor and Mellin decompositions yield the Gamow and real-energy scattering data, respectively, as summarized in \eqref{eq:Taylor-Mellin-distinction-sec3-rev}.

\subsection{Duality of the two stationary bridges}
\label{subsec:duality-stationary-bridges-v2}

In unit-frequency notation, \eqref{eq:UC-main-maps-sec4-v2} and \eqref{eq:quarter-rotations-sec2-rev} exhibit the crossed roles of the elliptic and hyperbolic generators:
\begin{center}
\begin{tabular}{lll}
\toprule
 & FP--HO bridge & FP--IHO bridge \\
\midrule
flow & complexified hyperbolic & real elliptic \\
generator & \(H_-\) & \(H_+\) \\
parameter & \(i\pi/4\) & \(\pm\pi/4\) \\
operator mapped & \(2iD\) & \(2D\) \\
target & \(H_+\) & \(\mp H_-\) \\
character & nonunitary similarity & unitary metaplectic rotation \\
\bottomrule
\end{tabular}
\end{center}
Thus the elliptic oscillator is reached through a complexified hyperbolic flow, while the hyperbolic inverted oscillator is reached through a real elliptic rotation.

The fractional-Fourier interpretation is a useful by-product of this picture.  The ordinary Fourier transform corresponds to a metaplectic rotation by \(\pi/2\), whereas the FP--IHO bridge uses \(\pm\pi/4\).  The classical canonical rotations through \(\pm\pi/4\) have order eight; at the metaplectic level the eighth power carries the usual central sign unless the oscillator zero-point phase is removed.  This classical order-eight structure is the real-rotation analogue of the order-eight property of the complex Cayley matrix entering the FP--HO bridge; the convention-dependent details are collected in Appendix~\ref{app:order-eight}.

The same crossed relation is reflected in Hermiticity. The Cayley exponential \eqref{eq:Cayley-operator-Hminus-sec4-v2} can also be viewed as evolution at real parameter \(\pi/4\) generated by the anti-Hermitian operator \(iH_-\); its similarity action maps the anti-Hermitian generator \(2iD\) to the Hermitian compact generator \(H_+\). By contrast, the real rotations \eqref{eq:quarter-rotations-sec2-rev}, generated by the Hermitian \(H_+\), relate Hermitian dilation and IHO generators.

\subsection{Relation with the analytic continuation \texorpdfstring{\(\omega\to\pm i\Omega\)}{omega to plus/minus i Omega}}
\label{subsec:omega-to-iOmega-v2}

There is also an indirect route, FP \(\to\) HO \(\to\) IHO.  The first step is the FP--HO bridge; the second is the analytic continuation \(\omega\to\pm i\Omega\).  At the level of Hamiltonians only \(\omega^2\to-\Omega^2\) is needed, but at the level of states the two signs matter because
\begin{equation}
E_n^{\rm HO}=\hbar\omega\left(n+\frac12\right)
\quad\longrightarrow\quad
E_n^{\pm}=\pm i\hbar\Omega\left(n+\frac12\right).
\label{eq:HO-to-IHO-Gamow-energies-sec4-v2}
\end{equation}
This is the analytic-continuation counterpart of keeping both real rotations \(\theta=\pm\pi/4\) in the direct FP--IHO bridge.

The continuation of the discrete oscillator spectrum gives the locations of the Gamow poles.  It does not by itself give the full real-energy IHO scattering amplitudes.  Those require connection coefficients: Fourier--Mellin coefficients in the direct FP--IHO route, or Weber-function connection formulae in the indirect FP--HO--IHO route.

The following sections develop the light-cone, scattering and
time-dependent realizations of these bridges.

\section{Light-cone variables, boundary-value prescriptions, and Gamow states}
\label{sec:light-cone-gamow-revised}

The FP--IHO bridge sends the dilation generator to the IHO Hamiltonian by the real rotations through \(\pm\pi/4\) discussed in Section~\ref{sec:conformal-bridges-revised-improved}.  We now diagonalize the hyperbolic generator itself.  This is most transparent in the light-cone variables adapted to the stable and unstable directions of the saddle.  They display at once the Mellin character of the real-energy spectrum, the role of the branch prescriptions at the separatrix, and the two complex Gamow ladders
\begin{equation}
E_n^{\pm}=\pm i\hbar\Omega\left(n+\frac12\right),
\qquad n=0,1,2,\ldots .
\label{eq:sec5-rev-Gamow-energies}
\end{equation}
The extraction of the real-energy transmission and reflection amplitudes from these light-cone data is left to Section~\ref{sec:scattering-Fourier-Mellin-revised-v2}.

\subsection{Light-cone form of the IHO Hamiltonian}
\label{subsec:sec5-rev-light-cone}

With the frequency-scaled canonical variables \(Q=\sqrt{\Omega}\,q\), \(P=p/\sqrt{\Omega}\), the IHO Hamiltonian is
\begin{equation}
H_{\rm IHO}=\frac{\Omega}{2}(P^2-Q^2).
\label{eq:sec5-rev-HIHO-QP}
\end{equation}
We use
\begin{equation}
u_+=\frac{P+Q}{\sqrt2},
\qquad
u_- =\frac{P-Q}{\sqrt2},
\qquad [u_+,u_-]=i\hbar,
\label{eq:sec5-rev-u-def}
\end{equation}
so that
\begin{equation}
H_{\rm IHO}=\frac{\Omega}{2}(u_+u_-+u_-u_+).
\label{eq:sec5-rev-HIHO-u}
\end{equation}
Classically this reduces to \(H_{\rm IHO}=\Omega u_+u_-\).  The symmetrized quantum form in \eqref{eq:sec5-rev-HIHO-u} is what produces the half-density shift below.

The Heisenberg equations give \(\dot u_+=\Omega u_+\), \(\dot u_-=-\Omega u_-\), hence
\begin{equation}
u_+(t)=e^{\Omega t}u_+(0),
\qquad
u_-(t)=e^{-\Omega t}u_-(0).
\label{eq:sec5-rev-boost-flow}
\end{equation}
Thus the \(u_+\)-direction, i.e. the separatrix \(u_-=0\), is unstable/outgoing for positive time, while the \(u_-\)-direction, i.e. the separatrix \(u_+=0\), is stable/incoming.  This is the convention used in the phase-space figure of 
Section~\ref{sec:phase-space-wigner}.

\subsection{Mellin eigenfunctions and boundary values}
\label{subsec:sec5-rev-Mellin-i0}

In the \(u_+\)-polarization, \(u_+=u\), \(u_-=-i\hbar d/du\), and \eqref{eq:sec5-rev-HIHO-u} becomes
\begin{equation}
H_{\rm IHO}=-i\hbar\Omega\left(u\frac{d}{du}+\frac12\right).
\label{eq:sec5-rev-H-u-rep}
\end{equation}
Therefore \(H_{\rm IHO}\psi_E=E\psi_E\) has local solutions
\begin{equation}
\psi_E(u)\sim u^\lambda,
\qquad
\lambda=-\frac12+i\epsilon,
\qquad
\epsilon=\frac{E}{\hbar\Omega}.
\label{eq:sec5-rev-lambda}
\end{equation}
These are multiplicative, or Mellin, characters: \((\alpha u)^\lambda=\alpha^\lambda u^\lambda\) for \(\alpha>0\).  This is the hyperbolic counterpart of the additive Fourier characters \(e^{ikq}\) of the free particle.

Although the equation is first order on each open half-line, the singular point \(u=0\) allows two independent distributional solutions on the full line.  A convenient real-energy basis is
\begin{equation}
\chi_{E,+}(u)=\Theta(u)u^{-1/2+iE/(\hbar\Omega)},
\qquad
\chi_{E,-}(u)=\Theta(-u)(-u)^{-1/2+iE/(\hbar\Omega)},
\label{eq:sec5-rev-halfline-basis}
\end{equation}
where \(\Theta\) denotes the Heaviside step function.
Equivalently one may use the boundary values \((u\pm i0)^\lambda\).  For \(u>0\) they coincide with \(u^\lambda\), while for \(u<0\)
\begin{equation}
(u+i0)^\lambda=e^{i\pi\lambda}(-u)^\lambda,
\qquad
(u-i0)^\lambda=e^{-i\pi\lambda}(-u)^\lambda.
\label{eq:sec5-rev-i0-negative}
\end{equation}
Since \(e^{\pm i\pi\lambda}=e^{\mp i\pi/2}e^{\mp\pi\epsilon}\), the branch choice already contains the exponential factors which later reappear in the Fourier--Mellin connection coefficients.  The normalized coefficients \(T(E)\) and \(R(E)\), however, require the flux-normalized calculation of Section~\ref{sec:scattering-Fourier-Mellin-revised-v2}.

The point \(u=0\) is therefore both a classical separatrix and a quantum branch point.  The prescription \((u\pm i0)^\lambda\) fixes how the solution is analytically continued around this channel boundary.  This is the light-cone analogue of the WKB choice of branch and contour around turning-point branch points.  In the stable oscillator the analogous global phase information appears through the Maslov correction, producing the familiar shift \(n+1/2\).  For the IHO the same half-density origin is seen locally in \eqref{eq:sec5-rev-H-u-rep}, while the branch prescription controls the hyperbolic scattering and resonant sectors.

\subsection{Gamow families}
\label{subsec:sec5-rev-Gamow}

The two Gamow families follow from boundary-value evaluations and
residues of the homogeneous distributions. For related treatments of
these resonance distributions, see
Refs.~\cite{Chruscinski2003,Chruscinski2004}.
At \(\lambda=n\), the boundary values \((u\pm i0)^\lambda\)
coincide with \(u^n\). In the \(u_+\)-polarization these monomials satisfy
\begin{equation}
H_{\rm IHO}u^n=-i\hbar\Omega\left(n+\frac12\right)u^n,
\qquad n=0,1,2,\ldots .
\label{eq:sec5-rev-Gamow-minus}
\end{equation}
With the time factor \(\exp(-iEt/\hbar)\), this family has quantized
amplitude decay rates \(\gamma_n=\Omega(n+1/2)\) for \(t>0\).
Gamow states are generalized states, not normalizable probability
states.

The time-reversed family is represented, in the same polarization, by
delta derivatives obtained from the meromorphically continued
half-line powers:
\[
\operatorname*{Res}_{\lambda=-n-1}\bigl[\Theta(u)u^\lambda\bigr]
=\frac{(-1)^n}{n!}\delta^{(n)}(u).
\]
Equivalently, these supported distributions are proportional to the
jump \((u+i0)^{-n-1}-(u-i0)^{-n-1}\), rather than to either boundary
value separately. Using
\(u(d/du)\delta^{(n)}(u)=-(n+1)\delta^{(n)}(u)\), one obtains
\begin{equation}
H_{\rm IHO}\delta^{(n)}(u)=+i\hbar\Omega\left(n+\frac12\right)\delta^{(n)}(u).
\label{eq:sec5-rev-Gamow-plus-supported}
\end{equation}
Together these two families give precisely
 \eqref{eq:sec5-rev-Gamow-energies}.  
 The signs distinguish resonant and anti-resonant boundary conditions: one family decays for positive time and the other for negative time, or equivalently after reversing the arrow of time.

The apparent asymmetry between polynomials and supported distributions is polarization-dependent.  In the conjugate \(u_-\)-representation, where \(u_-=v\), \(u_+=i\hbar d/dv\), the Hamiltonian reads
\begin{equation}
H_{\rm IHO}=+i\hbar\Omega\left(v\frac{d}{dv}+\frac12\right),
\label{eq:sec5-rev-H-v-rep}
\end{equation}
so that \(v^n\) represents the upper family, while the lower family is supported at \(v=0\).  Thus the two light-cone polarizations exchange the polynomial descriptions of the two Gamow ladders.

The stationary bridge identifies these families with \(\Gamma_n^{(\pm)}=R_{\pm\pi/4}q^n\), whose eigenvalue equations are given in \eqref{eq:Gamow-families-sec4-v2}. Thus the two IHO Gamow towers arise from the same FP threshold module that supplies the HO bound states.

\subsection{Linear ladder operators and real-energy states}
\label{subsec:sec5-rev-ladders-scattering-prep}

The ladder relations \eqref{eq:u-ladder-sec2-rev} imply that, if \(H_{\rm IHO}\psi_E=E\psi_E\), then \(u_+\psi_E\) has energy \(E-i\hbar\Omega\), whereas \(u_-\psi_E\) has energy \(E+i\hbar\Omega\).  The quadratic combinations \(u_+^2/2\), \(u_-^2/2\), and \((u_+u_-+u_-u_+)/2\) form the same \(\mathfrak{sl}(2,\mathbb R)\) algebra, while the linear operators extend it to the corresponding
 \(\mathfrak{osp}(1|2)\) structure.

It is important to distinguish the real-energy and Gamow uses of the same homogeneous distributions.  For \(E\in\mathbb R\), the half-line states \eqref{eq:sec5-rev-halfline-basis} and the boundary values \((u\pm i0)^{-1/2+iE/(\hbar\Omega)}\) belong to the generalized real-energy scattering basis of the self-adjoint IHO Hamiltonian.  At the exceptional values \eqref{eq:sec5-rev-Gamow-energies}, the
polynomial boundary values and the supported residues give the
resonant/anti-resonant Gamow states. Thus the scattering and resonance
data arise from the same light-cone homogeneous distributions, with
evaluation and residue extraction distinguished.

To obtain the physical scattering amplitudes one must compare the two conjugate light-cone polarizations.  Since \([u_+,u_-]=i\hbar\), this comparison is a Fourier transform.  Applied to half-line Mellin distributions, it produces Gamma functions and the exponential connection factors which, after flux normalization, give the IHO transmission and reflection coefficients.  This is the subject of Section~\ref{sec:scattering-Fourier-Mellin-revised-v2}.

\section{Scattering coefficients as Fourier--Mellin connection data}
\label{sec:scattering-Fourier-Mellin-revised-v2}

We consider the standard self-adjoint IHO Hamiltonian on the whole coordinate line,
with real energy and the usual two-channel scattering boundary conditions. In the
light-cone variables the stationary problem becomes the spectral problem of a
dilation generator. Its incoming and outgoing channels are naturally described by
half-line Mellin bases, and the Fourier transform between these bases gives the
fixed-energy connection matrix~\cite{Barton1990,BalazsVoros1990}.

The inverted parabolic potential does not approach a constant at spatial infinity.
The coordinate-space asymptotic channels are therefore flux-normalized Weber waves,
with quadratic and logarithmic phases, rather than plane waves. This does not
prevent a scattering interpretation: Appendix~\ref{app:weber-iho} identifies the
Mellin connection coefficients with the ratios of outgoing to incoming wave
amplitudes in that coordinate-space problem. Transmission at negative energy is
defined by the nonzero current beyond the classically forbidden interval, not by
a classical trajectory crossing a phase-space separatrix. Section~\ref{sec:phase-space-wigner}
explains the corresponding Wigner-function description.

This should be distinguished from the Gamow construction of 
Section~\ref{sec:light-cone-gamow-revised}.  Gamow states are pole and residue data of the analytically continued problem; the physical transmission and reflection amplitudes are real-energy connection coefficients.  Both are encoded in the same Fourier--Mellin structure, but in different regimes.

\subsection{Incoming and outgoing Mellin bases}
\label{subsec:sec6-rev-bases-v2}

We use the light-cone convention \([u_+,u_-]=i\hbar\) and the dilation
form \eqref{eq:sec5-rev-HIHO-u} of the IHO Hamiltonian.
For positive time, \(u_-\) labels the stable/incoming direction, while \(u_+\) labels the unstable/outgoing direction.  In the outgoing polarization, \(u_+=u\), one has \(H_{\rm IHO}=-i\hbar\Omega(u\,d/du+1/2)\), and hence
\begin{equation}
\chi^{\rm out}_{E,\pm}(u)=N_E\Theta(\pm u)(\pm u)^{-1/2+i\epsilon},
\qquad \epsilon=\frac{E}{\hbar\Omega} .
\label{eq:sec6-v2-out-basis}
\end{equation}
Here \((\pm u)^\lambda\) means a positive argument on the corresponding half-line.  In the incoming polarization, \(u_-=v\), the sign of the dilation generator is reversed, so that
\begin{equation}
\chi^{\rm in}_{E,\pm}(v)=N_E\Theta(\pm v)(\pm v)^{-1/2-i\epsilon}.
\label{eq:sec6-v2-in-basis}
\end{equation}
With \(N_E=(2\pi\hbar\Omega)^{-1/2}\), both the outgoing states
\eqref{eq:sec6-v2-out-basis} and the incoming states \eqref{eq:sec6-v2-in-basis}
are delta-normalized in energy:
\[
\langle E,\sigma;a|E',\sigma';a\rangle
=\delta_{\sigma\sigma'}\delta(E-E'),\qquad \sigma,\sigma'=\pm,\quad a=\mathrm{in},\mathrm{out}.
\]
Indeed, \(\int_0^\infty u^{-1-i\epsilon+i\epsilon'}du
=2\pi\delta(\epsilon'-\epsilon)\); the reversed exponent in the incoming
basis gives the same delta distribution. Powers of a dimensionful variable
are understood with a fixed reference scale in their logarithmic phase.

The two polarizations are related by the Fourier kernel
\begin{equation}
\langle u_+|u_-\rangle=(2\pi\hbar)^{-1/2}
\exp\left(\frac{i}{\hbar}u_+u_-\right).
\label{eq:sec6-v2-Fourier-kernel}
\end{equation}
An incoming state supported on \(v=u_->0\) decomposes into outgoing
\(u=u_+>0\) (transmission) and \(u<0\) (reflection) channels.
Starting from \(v<0\) gives the second column of the scattering matrix
\eqref{eq:sec6-v2-Smatrix}.

\subsection{Connection amplitudes}
\label{subsec:sec6-rev-amplitudes-v2}

Consider an incoming state supported on \(v=u_->0\).  Its outgoing wave in the \(u=u_+\) representation is
\begin{equation}
\Psi^{\rm out}_{E,+}(u)=\frac{N_E}{\sqrt{2\pi\hbar}}
\int_0^\infty \exp\left(\frac{iuv}{\hbar}\right)v^{-1/2-i\epsilon}\,dv .
\label{eq:sec6-v2-halfline-transform}
\end{equation}
Putting \(\mu=1/2-i\epsilon\), the regularized Mellin integrals
\begin{equation}
\int_0^\infty x^{\mu-1}e^{\pm isx}\,dx
=e^{\pm i\pi\mu/2}\Gamma(\mu)s^{-\mu},
\qquad s>0,
\label{eq:sec6-v2-Mellin-integrals}
\end{equation}
are the integral form of the \(i0\) prescription discussed above.  For \(u>0\) the phase is
\(\exp(iuv/\hbar)\), so one uses the upper sign in \eqref{eq:sec6-v2-Mellin-integrals}, with \(s=u/\hbar\).  For
\(u<0\), one writes \(u=-|u|\), so the phase becomes \(\exp(-i|u|v/\hbar)\), and one uses
the lower sign, with \(s=|u|/\hbar\).
 This gives
\begin{equation}
\Psi^{\rm out}_{E,+}=t(E)\chi^{\rm out}_{E,+}+r(E)\chi^{\rm out}_{E,-}.
\label{eq:sec6-v2-connection}
\end{equation}

To avoid assigning a logarithm to a dimensionful quantity, it is preferable to introduce dimensionless light-cone variables, for instance \(\xi=u/\sqrt{\hbar}\) and \(\eta=v/\sqrt{\hbar}\), or equivalently to insert an arbitrary reference scale in the Mellin powers.  Such a choice changes only a common phase.  In a dimensionless convention the amplitudes may be written as
\begin{equation}
\begin{aligned}
t(E)&=\frac{e^{i\varphi_0(E)}}{\sqrt{2\pi}}\,
\Gamma\left(\frac12-i\epsilon\right)
\exp\left[\frac{i\pi}{2}\left(\frac12-i\epsilon\right)\right],\\
r(E)&=\frac{e^{i\varphi_0(E)}}{\sqrt{2\pi}}\,
\Gamma\left(\frac12-i\epsilon\right)
\exp\left[-\frac{i\pi}{2}\left(\frac12-i\epsilon\right)\right],
\end{aligned}
\label{eq:sec6-v2-tr-amplitudes}
\end{equation}
where \(\varphi_0(E)\) is a convention-dependent real phase. Within this channel convention, their ratio is independent of \(\varphi_0(E)\):
\begin{equation}
\frac{r(E)}{t(E)}=-i\,e^{-\pi\epsilon}.
\label{eq:sec6-v2-relative-phase}
\end{equation}
Using \(|\Gamma(1/2-i\epsilon)|^2=\pi/\cosh(\pi\epsilon)\), one obtains
\begin{equation}
T(E)=|t(E)|^2=\frac{1}{1+e^{-2\pi E/(\hbar\Omega)}},
\qquad
R(E)=|r(E)|^2=\frac{1}{1+e^{2\pi E/(\hbar\Omega)}}.
\label{eq:sec6-v2-TR}
\end{equation}
The transmission probability is the parabolic-barrier result of
Kemble~\cite{Kemble1935}; see also Landau and
Lifshitz~\cite[Sec.~50, Problem~4, p.~184]{LandauLifshitzQM1977}.
Thus \(T(E)+R(E)=1\), and at the top of the barrier \(T(0)=R(0)=1/2\).  Equivalently,
\begin{equation}
T(E)=\frac12\left(1+\tanh\frac{\pi E}{\hbar\Omega}\right),
\qquad
R(E)=\frac12\left(1-\tanh\frac{\pi E}{\hbar\Omega}\right).
\label{eq:sec6-v2-TR-tanh}
\end{equation}
For a state incident from the left, with no incoming wave from the right,
Appendix~\ref{app:weber-iho} shows explicitly that
\[
T(E)=\frac{j_{\rm tr}}{j_{\rm in}},\qquad
R(E)=\frac{|j_{\rm ref}|}{j_{\rm in}}.
\]
In particular, for \(E<0\) the transmitted current is nonzero beyond the
forbidden interval \(|q|<\sqrt{-2E}/\Omega\) (unit mass). The same stationary
state has energy \(E\) on both sides. Neither a change of energy nor a reversal
of time is required. The crossover on the scale \(\hbar\Omega\) is thus already
a fixed-energy quantum effect, rather than a consequence of averaging a packet
over different energies.

It is also useful to note the symmetry
\[
T(-E)=R(E),\qquad R(-E)=T(E).
\]
This identity compares transmission and reflection at two different stationary
energies; it is not an energy-changing transition.

To specify the channel convention, let \(a_L,a_R\) denote incidence from the
left and right and \(b_L,b_R\) denote exit to the left and right. The incoming
signs \((u_->0,u_-<0)\) correspond to \((a_L,a_R)\), whereas the outgoing signs
\((u_+>0,u_+<0)\) correspond to \((b_R,b_L)\). In this sign ordering,
\begin{equation}
\begin{pmatrix}b_R\\b_L\end{pmatrix}
=S(E)\begin{pmatrix}a_L\\a_R\end{pmatrix},\qquad
S(E)=\begin{pmatrix}t(E)&r(E)\\r(E)&t(E)\end{pmatrix},\qquad
S(E)^\dagger S(E)=1.
\label{eq:sec6-v2-Smatrix}
\end{equation}
Unitarity follows from \eqref{eq:sec6-v2-TR} and
\(t^*r+r^*t=0\), implied by \eqref{eq:sec6-v2-relative-phase}.
Ordering both the incoming and outgoing spatial channels as left, right
gives
\[
\begin{pmatrix}b_L\\b_R\end{pmatrix}
=S_{LR}\begin{pmatrix}a_L\\a_R\end{pmatrix}.
\]
For reciprocal, flux-conserving scattering in flux-normalized channels,
the general form is
\(S_{LR}=\left(\begin{smallmatrix}r_L&t\\t&r_R\end{smallmatrix}\right)\),
where \(r_L\) and \(r_R\) are the reflection amplitudes for incidence
from the left and right, respectively. Unitarity implies
\(r_R=-r_L^*t/t^*\) for \(t\ne0\).
For the parity-symmetric IHO, with our matching channel phases,
\(r_L=r_R=r\); indeed,
Eq.~\eqref{eq:sec6-v2-relative-phase} gives
\(-r^*t/t^*=r\). Thus
\(S_{LR}=\left(\begin{smallmatrix}r&t\\t&r\end{smallmatrix}\right)\),
obtained from Eq.~\eqref{eq:sec6-v2-Smatrix} by interchanging
the outgoing rows.

The Fourier transform is a change of
representation, not the physical time evolution through the barrier; its
interpretation as a scattering matrix follows from the asymptotic current
identification. The noncommuting operators \(u_+\) and \(u_-\) define two
channel bases, not four simultaneously sharp quantum phase-space labels.

\subsection{Weber form and Gamow poles}
\label{subsec:sec6-rev-Weber-poles-v2}

The same connection data appear in the coordinate representation.  The IHO stationary equation
\begin{equation}
-\frac{\hbar^2}{2}\psi''(q)-\frac12\Omega^2q^2\psi(q)=E\psi(q)
\label{eq:sec6-v2-IHO-q-eq}
\end{equation}
is transformed by \(z=e^{-i\pi/4}\sqrt{2\Omega/\hbar}\,q\) into Weber's equation,
\begin{equation}
y''(z)+\left(\nu+\frac12-\frac{z^2}{4}\right)y(z)=0,
\qquad
\nu=-\frac12+\frac{iE}{\hbar\Omega}.
\label{eq:sec6-v2-Weber}
\end{equation}
The parabolic-cylinder functions \(D_\nu(z)\) have Stokes-sector connection formulae whose relative coefficients contain the same factors \(\exp[\pm\pi E/(2\hbar\Omega)]\) as the 
Mellin calculation; see Appendix~\ref{app:weber-iho}.  
Thus the light-cone and Weber languages describe the same scattering matrix: the former exposes the dilation and separatrix structure, while the latter is the usual whole-line Schr\"odinger formulation.

In the fixed Mellin phase convention \(\varphi_0=0\), the Gamma factors in \eqref{eq:sec6-v2-tr-amplitudes} also display the resonant structure. The poles of \(\Gamma(1/2-iE/(\hbar\Omega))\) occur at
\begin{equation}
E=-i\hbar\Omega\left(n+\frac12\right),
\qquad n=0,1,2,\ldots,
\label{eq:sec6-v2-minus-poles}
\end{equation}
while the time-reversed amplitudes contain \(\Gamma(1/2+iE/(\hbar\Omega))\) and give the opposite poles.  These are the same two Gamow families found in Section~\ref{sec:light-cone-gamow-revised}.  They are not normalizable eigenvalues of a self-adjoint Hamiltonian; they are resonance and anti-resonance poles of the analytically continued connection problem.

\subsection{Direct and indirect routes through the triangle}
\label{subsec:sec6-rev-routes-v2}

The direct FP--IHO route makes the origin of \eqref{eq:sec6-v2-tr-amplitudes} especially clear.  A free-particle plane wave is an additive character.  After a rotation through \(\pm\pi/4\) it becomes a generalized eigenstate of one light-cone operator, for instance a light-cone coherent state of the form \(\exp(i\eta u_+/\hbar)\).  This is not an energy eigenstate of the IHO.  Its energy decomposition is a Mellin transform, and the Mellin coefficients are precisely of the Gamma-function type appearing in \eqref{eq:sec6-v2-tr-amplitudes}.  Schematically,
\begin{equation}
\text{FP plane wave}\longrightarrow
\text{IHO light-cone coherent state}\longrightarrow
\text{Mellin coefficients }t(E),r(E).
\label{eq:sec6-v2-direct-route}
\end{equation}

The indirect route passes through the oscillator.  The FP--HO bridge maps the same plane wave to a HO coherent state \(|\alpha\rangle\).  Its Fock expansion,
\begin{equation}
|\alpha\rangle=e^{-|\alpha|^2/2}\sum_{n=0}^{\infty}\frac{\alpha^n}{\sqrt{n!}}|n\rangle,
\label{eq:sec6-v2-HO-coherent-Fock}
\end{equation}
is adapted to the elliptic generator \(H_+\).  After \(\omega\to\pm i\Omega\), the oscillator eigenvalues continue to the two IHO Gamow families; hence this discrete expansion reveals the pole positions.  It does not, however, yet give the real-energy scattering amplitudes.

To extract those amplitudes one has to expand the same coherent state in the hyperbolic basis of \(H_-\).  If \(|E,\sigma\rangle_-\), \(\sigma=\pm\), are generalized eigenstates of the IHO generator normalized on the real energy axis, then
\begin{equation}
|\alpha\rangle=\sum_{\sigma=\pm}\int_{-\infty}^{\infty}C_\sigma(E;\alpha)|E,\sigma\rangle_-\,dE,
\qquad
C_\sigma(E;\alpha)={}_-\langle E,\sigma|\alpha\rangle .
\label{eq:sec6-v2-HO-hyperbolic-expansion}
\end{equation}
Here the subscript ``\(-\)'' indicates that the states belong to the
hyperbolic \(H_-\)-basis, while \(\sigma=\pm\) labels the two channel
sectors, equivalently the two half-line light-cone branches in the
Mellin representation.
In coordinate representation these coefficients are overlap integrals between the oscillator coherent-state wave function and the IHO Weber scattering states,
\begin{equation}
C_\sigma(E;\alpha)=\int_{-\infty}^{\infty}
\Psi^{(-)}_{E,\sigma}(q)^*\,\Psi^{(+)}_\alpha(q)\,dq,
\label{eq:sec6-v2-coherent-Weber-overlap}
\end{equation}
where \(\Psi^{(-)}_{E,\sigma}\) is a Weber scattering state in a specified
incoming or outgoing basis and \(\Psi^{(+)}_\alpha\) is the HO coherent-state
Gaussian. These overlaps are state-dependent spectral weights; they are not
themselves the universal amplitudes \(t(E),r(E)\). The Weber connection
formulae relate their incoming and outgoing versions through the matrix
\eqref{eq:sec6-v2-Smatrix}. A generic HO coherent state contains both incoming
channels. Only after selecting one incoming component is its pair of outgoing
coefficients proportional to the corresponding column of \(S(E)\).
Appendix~\ref{subsec:appB-coherent} gives this extraction explicitly.

The two expansions therefore have distinct roles:
\begin{equation}
\begin{array}{c|c}
\text{elliptic/Fock expansion} & \text{HO spectrum and, after }\omega\to\pm i\Omega,\text{ Gamow poles}\\
\text{hyperbolic }H_-\text{ expansion} & \text{spectral weights related by the real-energy }S\text{ matrix}
\end{array}
\label{eq:sec6-v2-expansion-table}
\end{equation}
The two signs \(\pm\pi/4\) and \(\omega\to\pm i\Omega\) should therefore remain visible.  For probabilities alone their effect may be hidden by absolute values.  For amplitudes, phases, pole half-planes, and the distinction between resonant and anti-resonant boundary conditions, the two signs are essential.

\subsection{Thermal-looking factor}
\label{subsec:sec6-rev-thermal-factor-v2}

The result \eqref{eq:sec6-v2-TR} may be written as a Fermi-function-like expression,
\begin{equation}
T(E)=\frac{1}{1+\exp[-E/(k_B T_{\rm eff})]},
\qquad k_BT_{\rm eff}=\frac{\hbar\Omega}{2\pi}.
\label{eq:sec6-v2-thermal}
\end{equation}
For the one-particle IHO this is not a statement about fermions or genuine thermality.  It reflects the factor \(
\exp[-2\pi E/(\hbar\Omega)]\) generated by the Fourier--Mellin, or equivalently Weber/Stokes, connection problem.  Related connection factors enter Rindler, Unruh and Hawking analyses as well as Schwinger pair production. Their interpretation, including any claim of thermality, depends on the field state, mode normalization and boundary conditions discussed in Section~\ref{sec:physical-applications-final-style-v2}.

The Fourier--Mellin and Weber constructions thus give the same
stationary connection data. We next turn to the evolution problem.

\section{Hyperbolic Cayley--Niederer transformation and the TDSE}
\label{sec:hyperbolic-CN-TDSE}

The stationary FP--IHO bridge relates the dilation generator to the inverted-oscillator Hamiltonian by a real metaplectic rotation through \(\pm\pi/4\).  We now pass to the time-dependent counterpart.  It maps solutions of the free time-dependent Schr\"odinger equation (TDSE) into solutions of the IHO TDSE by a hyperbolic projective change of time, a scale transformation of the coordinate, and the corresponding metaplectic phase.

As in the preceding sections, we use units with mass equal to one. Restoring the mass gives the kinetic term \(p^2/(2m)\) and multiplies both the quadratic potential and the chirp phase by \(m\). The free and IHO TDSEs are
\begin{equation}
  \ii\hbar\partial_t\psi_0=-\frac{\hbar^2}{2}\partial_x^2\psi_0,
  \qquad
  \ii\hbar\partial_\tau\psi_-
  =\left(-\frac{\hbar^2}{2}\partial_q^2-\frac{1}{2}\Omega^2q^2\right)\psi_- .
  \label{eq:free-and-iho-TDSE}
\end{equation}
The hyperbolic \CN{} transformation intertwining them is
\begin{equation}
  t=\frac{1}{\Omega}\tanh(\Omega\tau),
  \qquad
  x=\frac{q}{\rho(\tau)},
  \qquad
  \rho(\tau)=\cosh(\Omega\tau),
  \label{eq:hyperbolic-CN-map}
\end{equation}
with
\begin{equation}
  \psi_-(q,\tau)=\rho^{-1/2}
  \exp\left[\frac{\ii}{2\hbar}\frac{\dot\rho}{\rho}q^2\right]
  \psi_0\left(\frac{q}{\rho},t(\tau)\right).
  \label{eq:hyperbolic-CN-wavefunction}
\end{equation}
Here a dot denotes \(\dd/\dd\tau\), and \(\dot\rho/\rho=\Omega\tanh(\Omega\tau)\).
The prefactor is the unitary half-density factor.  Indeed, at fixed \(\tau\),
\begin{equation}
  x=\frac{q}{\rho(\tau)},\qquad \dd x=\frac{\dd q}{\rho(\tau)},
  \qquad
  \int_{\RR}|\psi_-(q,\tau)|^2\dd q
  =\int_{\RR}|\psi_0(x,t(\tau))|^2\dd x .
  \label{eq:norm-preservation-CN}
\end{equation}
The exponential in \eqref{eq:hyperbolic-CN-wavefunction} is the quadratic metaplectic chirp; it removes the mixed derivative produced by the time-dependent scaling \(x=q/\rho(\tau)\).

\subsection{Projective time and the hyperbolic chart}
\label{subsec:projective-time-hyperbolic-chart}

The time map in \eqref{eq:hyperbolic-CN-map} satisfies
\begin{equation}
  \frac{\dd t}{\dd\tau}=1-\Omega^2t^2=\frac{1}{\rho^2(\tau)},
  \qquad
  \{t;\tau\}=-2\Omega^2,
  \label{eq:tanh-schwarzian}
\end{equation}
where \(\{t;\tau\}\) is the Schwarzian derivative.  It maps the complete IHO time line \(\tau\in\RR\) onto the finite free-particle interval \(|t|<1/\Omega\).  The endpoints \(t=\pm1/\Omega\) are fixed points of a hyperbolic one-parameter subgroup of \(SL(2,\RR)\) acting on projective time \(\RP^1\); they are chart boundaries, not singularities of the IHO dynamics.

The complementary chart may be chosen as
\begin{equation}
  t=-\frac{1}{\Omega}\coth(\Omega\tau),
  \qquad
  \rho(\tau)=|\sinh(\Omega\tau)|,
  \label{eq:hyperbolic-coth-chart}
\end{equation}
on either branch \(\tau\ne0\), where \(\dd t/\dd\tau=\Omega^2t^2-1>0\). This exterior chart is related to the tanh chart by the projective inversion \(t\mapsto-1/(\Omega^2t)\). The same transformation law \eqref{eq:hyperbolic-CN-wavefunction} applies, and again \(-\ddot\rho/\rho=-\Omega^2\). The two charts therefore describe the same local IHO dynamics on different projective intervals.

This chart structure is not peculiar to the hyperbolic case.  The free-particle conformal transformations are globally defined only after the affine time line is completed to \(\RP^1\).  The ordinary FP--HO Niederer map, \(t=\omega^{-1}\tan(\omega\tau)\), also covers only one affine chart at a time; crossing the points where \(\cos(\omega\tau)=0\) requires a chart change and, quantum mechanically, the corresponding metaplectic/Maslov phase bookkeeping.  The IHO case replaces the elliptic caustic sequence by the two hyperbolic fixed points and the complementary tanh/coth charts.

The constant-Schwarzian normal forms are
\begin{center}
\begin{tabular}{c c c c}
\toprule
case & time map \(t=f(\tau)\) & Schwarzian & induced system \\
\midrule
FP & \(f(\tau)=\tau\) & \(0\) & free \\
HO & \(f(\tau)=\omega^{-1}\tan(\omega\tau)\) & \(2\omega^2\) & \(H_+\) \\
IHO & \(f(\tau)=\Omega^{-1}\tanh(\Omega\tau)\) & \(-2\Omega^2\) & \(H_-\) \\
\bottomrule
\end{tabular}
\end{center}
Thus the sign of the Schwarzian is the TDSE manifestation of the elliptic versus hyperbolic distinction.

\subsection{General lens form and Ermakov--Niederer viewpoint}
\label{subsec:general-lens-form}

Equations \eqref{eq:hyperbolic-CN-map}--\eqref{eq:hyperbolic-CN-wavefunction} are a special case of the standard lens transformation.  Let \(t=f(\tau)\), \(\sigma(\tau)=\dot f(\tau)>0\), and set \(x=\sqrt{\sigma}\,q\).  If \(\psi_0(x,t)\) solves the free TDSE, then
\begin{equation}
  \psi(q,\tau)=\sigma^{1/4}
  \exp\left[-\frac{\ii}{4\hbar}\frac{\dot\sigma}{\sigma}q^2\right]
  \psi_0\bigl(\sqrt{\sigma}\,q,f(\tau)\bigr)
  \label{eq:general-lens-transform}
\end{equation}
solves
\begin{equation}
  \ii\hbar\partial_\tau\psi
  =\left[-\frac{\hbar^2}{2}\partial_q^2
  +\frac{1}{4}\{f;\tau\}q^2\right]\psi .
  \label{eq:general-lens-TDSE}
\end{equation}
Equivalently, with \(\sigma=\rho^{-2}\), one recovers \eqref{eq:hyperbolic-CN-wavefunction} and
\begin{equation}
  \omega_{\rm eff}^2(\tau)=\frac{1}{2}\{f;\tau\}=-\frac{\ddot\rho}{\rho} .
  \label{eq:effective-frequency}
\end{equation}
The choices \(\rho=1\), \(\rho=\cos(\omega\tau)\), and \(\rho=\cosh(\Omega\tau)\) give, respectively, the FP, HO, and IHO cases.

This is the elementary constant-frequency sector of the broader Ermakov--Niederer construction. For a general time-dependent quadratic Hamiltonian, the same metaplectic factorization is controlled by a scale function satisfying an associated linear or Ermakov equation.

\subsection{Action and constraint form}
\label{subsec:action-constraint-form}

The same map has a simple action-level interpretation.  From \(x=q/\rho\) and \(\dd t=\rho^{-2}\dd\tau\),
\begin{equation}
  \frac{1}{2}\left(\frac{\dd x}{\dd t}\right)^2\dd t
  =\left[\frac{1}{2}\dot q^{\,2}+\frac{1}{2}\frac{\ddot\rho}{\rho}q^2\right]\dd\tau
  -\dd\left(\frac{1}{2}\frac{\dot\rho}{\rho}q^2\right).
  \label{eq:action-lens-identity}
\end{equation}
For \(\rho=\cosh(\Omega\tau)\), the bulk term is the IHO Lagrangian \(L_- = \dot q^{\,2}/2+\Omega^2q^2/2\). The free and IHO actions differ by the boundary term in \eqref{eq:action-lens-identity}.  In the reparametrization-invariant formulation the free constraint \(C_0=p_t+p_x^2/2\approx0\) is mapped to
\begin{equation}
  C_- = \rho^{-2}C_0 = p_\tau + \frac12 p_q^2-\frac12\Omega^2q^2\approx0.
  \label{eq:IHO-constraint}
\end{equation}
The momentum shift, \(p_x=\rho p_q-\dot\rho q\), is lifted quantum 
mechanically by the chirp factor in \eqref{eq:hyperbolic-CN-wavefunction},
which is the metaplectic, quantum counterpart of the boundary term in \eqref{eq:action-lens-identity}.

\subsection{Wave packets, propagator, and phase-space meaning}
\label{subsec:wave-packets-propagator}

The hyperbolic \CN{} map sends free Gaussian packets into IHO Gaussian packets, whose centers and covariance ellipses follow the symplectic flow described in Section~\ref{sec:phase-space-wigner}.

The propagator follows from the free kernel by the same transformation.  For \(\tau\ne\tau'\),
\begin{equation}
\begin{split}
  K_-(q,\tau;q',\tau')
  ={}&\left(\frac{\Omega}{2\pi\ii\hbar\sinh\Omega(\tau-\tau')}\right)^{1/2} \\
  &\times\exp\left\{\frac{\ii\Omega}{2\hbar\sinh\Omega(\tau-\tau')}
  \left[(q^2+q'^2)\cosh\Omega(\tau-\tau')-2qq'\right]\right\} .
\end{split}
\label{eq:IHO-propagator-sec7}
\end{equation}
This is the hyperbolic counterpart of the oscillator Mehler kernel and is formally obtained from it by \(\omega\to\pm\ii\Omega\).  Since the propagator is a matrix element of the evolution operator, the apparent sign choice in \(\Omega\) is only a convention: the expression \eqref{eq:IHO-propagator-sec7} is invariant under \(\Omega\to-\Omega\), with the square-root branch chosen consistently.

\subsection{Relation to the stationary bridge}
\label{subsec:TDSE-SSE-bridge-comparison}

The stationary bridge organizes generalized spectral states and their
connection coefficients, whereas the TDSE bridge relates wave packets
and evolution kernels. Both are controlled by the projective/metaplectic
structure \(SL(2,\RR)\simeq Sp(2,\RR)\).
Near a smooth barrier top, \(V(q)=V_0-\Omega^2q^2/2+O(q^3)\), the
hyperbolic \CN{} map gives a local construction of IHO packet evolution
from free evolution. 
The TDSE evolution encodes the same transmission and reflection
amplitudes, which are extracted at fixed energy by asymptotic
channel matching in the Weber/Fourier--Mellin problem.

\section{Phase-space portrait and semiclassical interpretation}
\label{sec:phase-space-wigner}

The inverted oscillator is the elementary local model of hyperbolic motion
near a nondegenerate phase-space saddle. We use the scaled canonical variables
and light-cone coordinates of \eqref{eq:sec5-rev-HIHO-QP}--\eqref{eq:sec5-rev-HIHO-u}.
The classical symbol is \(H_{\rm cl}=\Omega(P^2-Q^2)/2=\Omega u_+u_-\),
and the flow \eqref{eq:sec5-rev-boost-flow} makes \(u_+\) unstable/outgoing
and \(u_-\) stable/incoming for positive time.

\subsection{Classical channel geometry}
\label{subsec:sec8-classical-channel-geometry}

The signs of \(u_-\) and \(u_+\) label the incoming and outgoing classical
branches. On \(H_{\rm cl}=E>0\) they agree: the upper and lower hyperbolae
in Fig.~\ref{fig:sec8-IHO-phase-space-wigner}, distinguished by the sign of
\(P\), describe transmission across the barrier. On \(H_{\rm cl}=E<0\)
the signs differ: the left and right hyperbolae describe reflection back to
the incident side. The zero-energy set consists of the unstable separatrix
\(P=Q\), where \(u_-=0\) and flow is outward, and the stable separatrix
\(P=-Q\), where \(u_+=0\) and flow is inward. No classical trajectory
crosses these separatrices.

The rescalings \(u_\pm(t)=e^{\pm\Omega t}u_\pm(0)\) have the form of a
Lorentz boost in null coordinates, with rapidity \(\Omega t\). This is an
analogy between hyperbolic flows, not ordinary spacetime motion, and does not
by itself determine scattering amplitudes. Nor do the four classical regions
give four simultaneously sharp quantum channel labels: the incoming and
outgoing sign projectors are functions of conjugate operators. As explained
below, a sharp-energy Wigner distribution need not be confined to the
classical curve \(H_{\rm cl}=E\)~\cite{BalazsVoros1990}.

\subsection{Wigner transport and Gaussian squeezing}
\label{subsec:sec8-Wigner-transport}

For a density operator \(\hat\rho\), define
\begin{equation}
  W_{\rho}(Q,P)=\frac{1}{2\pi\hbar}
  \int_{-\infty}^{+\infty} e^{-iPy/\hbar}
  \left\langle Q+\frac{y}{2}\right|\hat\rho\left|Q-\frac{y}{2}\right\rangle\,dy .
\label{eq:sec8-Wigner-rho}
\end{equation}
For \(\hat\rho=|\psi\rangle\langle\psi|\), the integrand contains
\(\psi(Q+y/2)\overline{\psi(Q-y/2)}\). The marginals
\(\int W_\rho\,dP=\langle Q|\hat\rho|Q\rangle\) and
\(\int W_\rho\,dQ=\langle P|\hat\rho|P\rangle\) are probability densities,
although \(W_\rho\) itself may be negative.

For the quadratic IHO Hamiltonian, the Moyal evolution equation reduces
exactly to the classical Liouville equation,
\begin{equation}
  \partial_t W+\Omega P\,\partial_Q W+\Omega Q\,\partial_P W=0.
\label{eq:sec8-Wigner-Liouville}
\end{equation}
Writing \(z=(Q,P)^T\), \(z(t)=M_{\rm IHO}(t)z(0)\), and \(W_0(z)=W(z,0)\),
its solution is
\begin{equation}
M_{\rm IHO}(t)=
\begin{pmatrix}
\cosh(\Omega t)&\sinh(\Omega t)\\
\sinh(\Omega t)&\cosh(\Omega t)
\end{pmatrix},\qquad
W(z,t)=W_0\bigl(M_{\rm IHO}^{-1}(t)z\bigr).
\label{eq:sec8-Wigner-flow}
\end{equation}
Equivalently,
\begin{equation}
  W(u_+,u_-;t)=W_0\left(e^{-\Omega t}u_+,e^{\Omega t}u_-\right).
\label{eq:sec8-Wigner-flow-u}
\end{equation}
This is the exact Egorov/metaplectic covariance of quadratic
dynamics~\cite{Folland1989,EgoTeor}.

An initially circular minimum-uncertainty Gaussian,
\(W_0=(\pi\hbar)^{-1}\exp[-(Q^2+P^2)/\hbar]\), therefore evolves as
\begin{equation}
  W_t(u_+,u_-)=\frac{1}{\pi\hbar}
  \exp\left[-\frac{e^{-2\Omega t}u_+^2+e^{2\Omega t}u_-^2}{\hbar}\right].
\label{eq:sec8-Wigner-evolved}
\end{equation}
Its widths obey \(\Delta u_\pm(t)=e^{\pm\Omega t}\Delta u_\pm(0)\):
the circular blob becomes an equal-area ellipse, stretched along \(u_+\)
and squeezed along \(u_-\), as in Fig.~\ref{fig:sec8-IHO-phase-space-wigner}.
For a displaced Gaussian, the center follows the classical trajectory and
the covariance ellipse follows the same symplectic flow. This Gaussian
example illustrates squeezing; the stationary current ratios are recovered
from scattering-state Wigner distributions in
Appendix~\ref{subsec:appB-wigner}.

\begin{figure}[t]
\centering
\begin{tikzpicture}[x=0.92cm,y=0.92cm,>=Stealth,line cap=round,line join=round]

\definecolor{myblue}{RGB}{35,85,255}
\definecolor{myorange}{RGB}{235,120,35}
\definecolor{myred}{RGB}{220,55,40}
\definecolor{mygreen}{RGB}{40,145,60}
\definecolor{mycyan}{RGB}{0,50,255}
\definecolor{mymagenta}{RGB}{255,0,255}

\def\rblob{0.34}
\def\aell{1.05}
\pgfmathsetmacro{\bell}{\rblob*\rblob/\aell}

\draw[->,thick] (-4.8,0) -- (5.0,0) node[right] {$Q$};
\draw[->,thick] (0,-3.9) -- (0,4.0) node[above] {$P$};

\draw[thick] (-3.5,-3.5) -- (3.5,3.5);
\draw[thick] (-3.5, 3.5) -- (3.5,-3.5);

\draw[->,thick] (-2.2,2.2) -- (-3.5,3.5);
\draw[->,thick] ( 2.2,2.2) -- ( 3.5,3.5);

\draw[->,black,line width=0.75pt] ( 0.95, 0.95) -- ( 1.55, 1.55);
\draw[->,black,line width=0.75pt] (-0.95,-0.95) -- (-1.55,-1.55);
\draw[->,black,line width=0.75pt] (-1.55, 1.55) -- (-0.95, 0.95);
\draw[->,black,line width=0.75pt] ( 1.55,-1.55) -- ( 0.95,-0.95);

\node at (-2.50,3.30) {$u_{-}>0$};
\node at ( 2.50,3.30) {$u_{+}>0$};

\draw[mygreen,thick,domain=-2.75:2.75,samples=120,smooth,variable=\x] plot (\x,{sqrt(\x*\x+0.95)});
\draw[mygreen,thick,domain=-2.75:2.75,samples=120,smooth,variable=\x] plot (\x,{sqrt(\x*\x+1.40)});
\draw[mygreen,thick,domain=-2.75:2.75,samples=120,smooth,variable=\x] plot (\x,{sqrt(\x*\x+1.85)});
\draw[myblue,thick,domain=-2.75:2.75,samples=120,smooth,variable=\x] plot (\x,{-sqrt(\x*\x+0.95)});
\draw[myblue,thick,domain=-2.75:2.75,samples=120,smooth,variable=\x] plot (\x,{-sqrt(\x*\x+1.40)});
\draw[myblue,thick,domain=-2.75:2.75,samples=120,smooth,variable=\x] plot (\x,{-sqrt(\x*\x+1.85)});
\draw[myred,thick,domain=-2.75:2.75,samples=120,smooth,variable=\y] plot ({sqrt(\y*\y+0.95)},\y);
\draw[myred,thick,domain=-2.75:2.75,samples=120,smooth,variable=\y] plot ({sqrt(\y*\y+1.40)},\y);
\draw[myred,thick,domain=-2.75:2.75,samples=120,smooth,variable=\y] plot ({sqrt(\y*\y+1.85)},\y);
\draw[myorange,thick,domain=-2.75:2.75,samples=120,smooth,variable=\y] plot ({-sqrt(\y*\y+0.95)},\y);
\draw[myorange,thick,domain=-2.75:2.75,samples=120,smooth,variable=\y] plot ({-sqrt(\y*\y+1.40)},\y);
\draw[myorange,thick,domain=-2.75:2.75,samples=120,smooth,variable=\y] plot ({-sqrt(\y*\y+1.85)},\y);

\foreach \c in {0.95,1.40,1.85}{
  \draw[->,mygreen,line width=0.6pt,domain=-0.95:-0.50,samples=30,smooth,variable=\x] plot (\x,{sqrt(\x*\x+\c)});
  \draw[->,mygreen,line width=0.6pt,domain=0.35:0.80,samples=30,smooth,variable=\x] plot (\x,{sqrt(\x*\x+\c)});
  \draw[->,myblue,line width=0.6pt,domain=0.95:0.50,samples=30,smooth,variable=\x] plot (\x,{-sqrt(\x*\x+\c)});
  \draw[->,myblue,line width=0.6pt,domain=-0.35:-0.80,samples=30,smooth,variable=\x] plot (\x,{-sqrt(\x*\x+\c)});
  \draw[->,myred,line width=0.6pt,domain=-0.95:-0.50,samples=30,smooth,variable=\y] plot ({sqrt(\y*\y+\c)},\y);
  \draw[->,myred,line width=0.6pt,domain=0.35:0.80,samples=30,smooth,variable=\y] plot ({sqrt(\y*\y+\c)},\y);
  \draw[->,myorange,line width=0.6pt,domain=0.95:0.50,samples=30,smooth,variable=\y] plot ({-sqrt(\y*\y+\c)},\y);
  \draw[->,myorange,line width=0.6pt,domain=-0.35:-0.80,samples=30,smooth,variable=\y] plot ({-sqrt(\y*\y+\c)},\y);
}

\filldraw[fill=mycyan,fill opacity=0.13,draw=myblue,thick] (0,0) circle (\rblob);
\begin{scope}[rotate=45]
  \filldraw[fill=mymagenta,fill opacity=0.16,draw=myorange,thick]
  (0,0) ellipse ({\aell} and {\bell});
\end{scope}
\draw[->,thick] (0.15,0.36) .. controls (-0.10,0.95) and (0.30,1.06) .. (0.58,0.70);

\node[mygreen]  at (0.78,2.65) {\large $H_{\rm cl}>0$};
\node[myblue]   at (0.78,-2.70) {\large $H_{\rm cl}>0$};
\node[myorange] at (-3.85,0.50) {\large $H_{\rm cl}<0$};
\node[myred]    at ( 3.85,0.50) {\large $H_{\rm cl}<0$};

\draw[mycyan,thin] (0.31,-0.11) -- (2.25,-0.75);
\node[mycyan,anchor=west] at (2.40,-0.75) {initial Wigner blob};
\node[mycyan,anchor=west] at (3.00,-1.13) {$(t=0)$};
\draw[mymagenta,thin] (-0.65,-0.70) -- (2.25,-1.50);
\node[mymagenta,anchor=west] at (2.40,-1.50) {evolved Wigner blob};
\node[mymagenta,anchor=west] at (3.05,-1.85) {$(t>0)$};

\end{tikzpicture}
\caption{Classical level sets of \(H_{\rm cl}=\Omega(P^2-Q^2)/2\) and
Gaussian-packet transport. Arrowheads at the ends of the axes indicate positive directions:
\(u_-\) toward the upper left and \(u_+\) toward the upper right. Arrows along
trajectories indicate forward time evolution; the black arrows nearer the
saddle point outward on \(P=Q\) and inward on \(P=-Q\). Upper/lower hyperbolae
transmit classically; left/right hyperbolae reflect. The Gaussian illustrates
stretching along \(u_+\) and squeezing along \(u_-\); it is not the Wigner
distribution of a stationary scattering eigenstate.}
\label{fig:sec8-IHO-phase-space-wigner}
\end{figure}

\subsection{Sharp quantum energy and phase-space support}
\label{subsec:sec8-smooth-crossover}

The classical step functions \(T_{\rm cl}(E)=\Theta(E)\) and
\(R_{\rm cl}(E)=\Theta(-E)\) become the smooth current ratios
\eqref{eq:sec6-v2-TR}--\eqref{eq:sec6-v2-TR-tanh}, with crossover scale
\(\hbar\Omega\) and \(T(0)=R(0)=1/2\). This is a fixed-energy quantum
effect, not a consequence of the energy spread of a packet.

To reconcile it with exact Liouville transport, distinguish the classical
symbol \(H_{\rm cl}\) from the quantum eigenvalue \(E\). The Wigner
distribution of a sharp-energy state satisfies both the transport equation
and the star-eigenvalue conditions
\begin{equation}
H_{\rm cl}\star W_E=E W_E=W_E\star H_{\rm cl},\qquad
(H_{\rm cl}-E)W_E+
\frac{\hbar^2\Omega}{8}(\partial_P^2-\partial_Q^2)W_E=0.
\label{eq:sec8-star-energy}
\end{equation}
Here \(\star\) denotes the standard Moyal product in the canonical
phase-space variables \((Q,P)\). The second relation is the real part of
the first. It is not the classical constraint \((H_{\rm cl}-E)W_E=0\);
sharp quantum energy therefore does not restrict support to \(H_{\rm cl}=E\).

For left incidence, put \(\xi=u_+/\sqrt\hbar\), \(\eta=u_-/\sqrt\hbar\),
and \(h=H_{\rm cl}/(\hbar\Omega)=\xi\eta\). In the flux convention of
Appendix~\ref{subsec:appB-wigner}, the dimensionless distribution is
\begin{equation}
\mathcal W_\epsilon(\xi,\eta)=\Theta(\eta)w_\epsilon(h),
\qquad \epsilon=\frac{E}{\hbar\Omega}.
\label{eq:sec8-Wigner-stationary}
\end{equation}
It vanishes for \(\eta<0\), but extends to \(h>0\) even when \(E<0\).
The positive-\(h\) contribution carries transmitted current, while the
right negative-\(h\) wedge, where \(\eta<0\), has no support. The nonlocal
Wigner transform is not a collection of particles with simultaneously
definite position and momentum. Balazs and Voros explain this support
property in Sec.~V and recover the currents in Sec.~VI and Appendix~E
of Ref.~\cite{BalazsVoros1990}; Appendix~\ref{subsec:appB-wigner} gives the
distribution and current integrals in our conventions. No classical
characteristic crosses a separatrix, and no quantum change \(E\to-E\)
is involved. The identity \(T(-E)=R(E)\) compares different stationary energies.

Phase-space horizons and sector assignments also enter
Refs.~\cite{GaztanagaKumarMarto2026,KumarMarto2026}, which propose a
direct-sum construction with geometric superselection sectors. This adds
assumptions about the quantum state space. Here the channels are two bases
of the standard whole-line IHO spectrum: \(\Theta(\hat u_+)\) and
\(\Theta(\hat u_-)\) each commute with IHO evolution but not with one
another. Their noncommutativity excludes jointly sharp incoming and outgoing
sign labels, not coordinate-space transmission. The classical obstruction
between negative-energy wedges is therefore not a prohibition of quantum
tunnelling in this scattering problem.

For normalizable packets, integration over a complete half-plane associated
with one linear quadrature gives a genuine probability, for example
\(\int_{u_+>0}W\,dQ\,dP=\langle\Theta(\hat u_+)\rangle\), even if \(W\)
is negative elsewhere. Arbitrary quadrant integrals need not be probabilities
of quantum projectors. For stationary delta-normalized states, transmission
and reflection are current ratios, not unnormalized phase-space areas.
The water-wave experiment of Rozenman et al.~\cite{Rozenman2024} uses free
propagation of tailored Weber profiles to expose the phase-space horizon
and its logarithmic phase singularity, and extracts the associated
Fermi--Dirac-type transmission/reflection factors. This classical wave
experiment illustrates the connection structure in a setting distinct
from one-particle IHO scattering.

Finally, the separatrix \(u_+=0\) is also the branch locus of the Mellin
eigenfunctions. The boundary values \((u\pm i0)^{-1/2+iE/(\hbar\Omega)}\)
and their negative-half-line phases are given in
\eqref{eq:sec5-rev-i0-negative}. As with branch choices in WKB theory, the
prescription fixes analytic connection data; it does not describe a
time-dependent trajectory crossing \(u=0\). The phase-space geometry
organizes the channels, while their boundary conditions and flux
normalization determine the observable \(T(E)\) and \(R(E)\).

\section{Coherent states, Perelomov states, and Bogoliubov transformations}
\label{sec:coherent-perelomov-bogoliubov}

The coherent-state viewpoint connects the generating functions of the
free-particle threshold module to oscillator states and IHO spectral data.
Three objects must be distinguished: generalized eigenstates of linear
light-cone operators, normalizable squeezed packets, and generalized
eigenstates of the IHO Hamiltonian. Only the last are stationary scattering
states; the detailed squeezed-state construction is collected in
Appendix~\ref{app:perelomov-packets}.

\subsection{Linear eigenstates and generating functions}
\label{subsec:sec9-FP-HO-Glauber}

Under the FP--HO bridge \eqref{eq:Somega-linear-map-sec4-v2}, the free plane
wave \(\psi_k(q)=e^{ikq}\) becomes an eigenstate of \(a^-\), with
\(\alpha(k)=i\hbar k/\sqrt{2\hbar\omega}\). The normalized Glauber
state~\cite{WallsMilburn1994} has the Fock expansion
\eqref{eq:sec6-v2-HO-coherent-Fock}.
This is the generating-function version of the Jordan-to-Fock map:
the Taylor coefficients of \(e^{ikq}\) at \(k=0\) are the threshold
monomials, whose bridge images form the oscillator tower.

For the IHO, the linear operator \(u_-\) instead has generalized eigenstates
\begin{equation}
 u_-\Phi_\eta=\eta\Phi_\eta,\qquad
 \Phi_\eta(u)=\frac{e^{i\eta u/\hbar}}{\sqrt{2\pi\hbar}},\qquad u=u_+.
\label{eq:sec9-IHO-linear-coherent-revised}
\end{equation}
Calling these light-cone coherent states refers only to the linear
eigenvalue equation. They are not normalizable and, for \(\eta\ne0\), are not eigenstates of
the dilation Hamiltonian \eqref{eq:sec5-rev-H-u-rep}.
For \(\eta\ne0\), their expansion in
the real-energy IHO basis is a Mellin expansion: applying
\eqref{eq:sec6-v2-Mellin-integrals} with \(\mu=1/2-iE/(\hbar\Omega)\)
and Fourier scale \(|\eta|/\hbar\) gives the same Gamma-function connection
data as in Section~\ref{sec:scattering-Fourier-Mellin-revised-v2}.
Their Taylor expansion, \(e^{i\eta u/\hbar}=\sum_{n\geq0}(i\eta/\hbar)^n u^n/n!\),
instead involves the Gamow polynomials, since
\(H_{\rm IHO}u^n=-i\hbar\Omega(n+1/2)u^n\). The Taylor and Mellin readings
thus probe resonant and real-energy sectors of the same generating function.

The IHO has no normalizable lowest-energy state on which to build an
ordinary Glauber vacuum. For example, the constant function annihilated
by \(u_-=-i\hbar\partial_u\) is the formal Gamow vector with energy
\(-i\hbar\Omega/2\), not a Hilbert-space vacuum. Likewise, \(E=0\) in
the scattering spectrum is characterized by the Mellin weight
\(|u|^{-1/2}\) on each half-line and \(T=R=1/2\), not by a normalizable
lowest-weight state.

\subsection{Normalizable packets and parity sectors}
\label{subsec:sec9-Perelomov-revised}

The quadratic \(\mathfrak{su}(1,1)\) generators preserve the even and odd
metaplectic sectors \(D^+_{1/4}\) and \(D^+_{3/4}\). Their Perelomov
states~\cite{Perelomov} are squeezed versions of \(|0\rangle\) and
\(|1\rangle\), respectively. The squeezed vacuum is a centered Gaussian;
the squeezed one-particle state is odd and non-Gaussian. A Heisenberg--Weyl
displacement of the squeezed vacuum gives a general pure Gaussian packet,
whose center and covariance evolve as in Section~\ref{sec:phase-space-wigner}.

The linear \(\mathfrak{osp}(1|2)\) generators \(a^\pm\), by contrast,
intertwine the parity sectors. A Glauber state with \(\alpha\ne0\)
contains both even and odd components; \(\alpha=0\) is the even vacuum
exception. These are different constructions within the same metaplectic
representation, not different spectra assigned to the IHO Hamiltonian.
Appendix~\ref{app:perelomov-packets} gives the explicit squeeze operators
and a compact classification. Normalizable packets have spectral envelopes;
the universal scattering matrix acts on those envelopes as described in
Appendix~\ref{subsec:appB-coherent}.

\subsection{Bogoliubov transformations and particle interpretation}
\label{subsec:sec9-Bogoliubov-revised}

For one oscillator mode a linear canonical transformation reads
\begin{equation}
 b^-=\alpha a^-+\beta a^+,\qquad b^+=(b^-)^\dagger,
 \qquad |\alpha|^2-|\beta|^2=1.
\label{eq:sec9-Bogoliubov-general-revised}
\end{equation}
With \(a^-=(Q+iP)/\sqrt{2\hbar}\), IHO Heisenberg evolution gives
\begin{equation}
 a^-(t)=a^-(0)\cosh(\Omega t)+i a^+(0)\sinh(\Omega t),
 \qquad a^+(t)=[a^-(t)]^\dagger .
\label{eq:sec9-a-Bogoliubov-time-revised}
\end{equation}
This is the operator form of the hyperbolic flow
\eqref{eq:sec5-rev-boost-flow}. Its field-theoretic particle interpretation
requires a choice of mode decompositions and vacua. If
\begin{equation}
 a^{\rm out}_j=\sum_k\left(\alpha_{jk}a^{\rm in}_k+
 \beta_{jk}a^{{\rm in}\dagger}_k\right),\qquad
 \langle0,{\rm in}|N^{\rm out}_j|0,{\rm in}\rangle=\sum_k|\beta_{jk}|^2,
\label{eq:sec9-QFT-Bogoliubov-revised}
\end{equation}
nonzero \(\beta\) describes particles relative to the out-modes in the
in-vacuum. The one-particle IHO relation \(T+R=1\) and the bosonic
condition \(|\alpha|^2-|\beta|^2=1\) have different physical normalizations,
despite the shared hyperbolic and analytic structure. This distinction
governs the applications below.

\section{Physical applications and interpretation}
\label{sec:physical-applications-final-style-v2}

The triangle organizes several established appearances of the IHO through
its relation to dilations, Mellin modes, and Weber connection data. We retain
the physical distinctions between a local saddle approximation, a
time-dependent change of representation, and a field-theoretic choice of
modes or state space.

\subsection{Local saddles and barrier scattering}
\label{subsec:sec10-local-normal-form-v2}

Near a nondegenerate potential maximum,
\begin{equation}
 V(x)=V(x_0)-\frac{m\Omega^2}{2}(x-x_0)^2+O((x-x_0)^3),
 \qquad \Omega^2=-V''(x_0)/m>0.
\label{eq:sec10-local-barrier-v2}
\end{equation}
Subtracting \(V(x_0)\) gives the IHO normal form, related to a dilation
generator by the real \(\pi/4\) rotation of Section~\ref{subsec:FP-IHO-bridge-v2}.
Its transmission and reflection probabilities are \eqref{eq:sec6-v2-TR}; their use for a
general barrier is a local approximation, whose range depends on the
neglected terms and on matching to the exterior potential.

In the quantum Hall problem, a strong perpendicular magnetic field separates
the fast Landau oscillator from slow guiding-center motion near a smooth
saddle~\cite{SubramanyanHegdeVishveshwaraBradlyn2021,Hegde2019}.
For \(V(x,y)=V_0+(U_y y^2-U_x x^2)/2\), \(U_x,U_y>0\), projection
to one Landau level gives
\begin{equation}
 H_{\rm gc}=\widetilde V_0+\frac12(U_yY^2-U_xX^2),\qquad
 [X,Y]=i\ell_B^2,\qquad \ell_B^2=\frac{\hbar c}{|eB|},
\label{eq:sec10-Hgc-v2}
\end{equation}
with a choice of orientation fixing the commutator sign. The Landau energy
and constant projection shifts are included in \(\widetilde V_0\).
Rescaling \(Q=(U_x/U_y)^{1/4}X/\ell_B\) and
\(P=(U_y/U_x)^{1/4}Y/\ell_B\) gives \([Q,P]=i\) and
\(H_{\rm gc}-\widetilde V_0=\ell_B^2\sqrt{U_xU_y}(P^2-Q^2)/2\).
The saddle contribution to point-contact transmission therefore has the
IHO form, with \(\hbar\Omega_{\rm gc}=\ell_B^2\sqrt{U_xU_y}\);
the embedding in the sample supplies the remaining boundary data.

Black-hole perturbations offer another barrier realization, distinct from
near-horizon kinematics~\cite{JaramilloLenziSopuerta2024}.
Their radial equations often have the form
\begin{equation}
 \left[-\frac{d^2}{dr_*^2}+V_{\rm eff}(r_*)\right]\Psi=\omega^2\Psi,
 \qquad
 V_{\rm eff}(r_*)=V_0-\frac{\gamma^2}{2}(r_*-r_{*0})^2+\cdots
\label{eq:sec10-RW-v2}
\end{equation}
near a maximum, with \(\gamma^2=-V_{\rm eff}''(r_{*0})>0\).
The local spectral parameter is \(\omega^2-V_0\). For Schwarzschild,
\(dr_*/dr=(1-2M/r)^{-1}\), and the barrier maximum is away from the
horizon, where the effective potential instead tends to zero. The IHO
approximation describes this maximum; the full greybody factors and
quasinormal spectra require the global potential and boundary conditions.
A related local instability occurs for particle motion near a horizon
under an external force~\cite{HashimotoTanahashi2017}.

\subsection{Field modes, horizons, and time-dependent representations}
\label{subsec:sec10-Rindler-v2}

For Schwinger production in a constant electric field, scalar modes obey,
in units \(\hbar=c=1\),
\begin{equation}
 \left[\frac{d^2}{dt^2}+(k+e\mathcal E t)^2+m_\perp^2\right]\phi_k(t)=0,
 \qquad m_\perp^2=m^2+k_\perp^2.
\label{eq:sec10-schwinger-mode-v2}
\end{equation}
An affine rescaling reduces this to the Weber equation~\cite{Schwinger1951,Dunne2004}.
The connection data now relate positive- and negative-frequency modes and
determine Bogoliubov coefficients, with the interpretation of
\eqref{eq:sec9-QFT-Bogoliubov-revised}, rather than spatial reflection and
transmission currents.

The Rindler, Unruh, and near-horizon Hawking settings display the same
Fourier--Mellin mechanism~\cite{Fulling1973,Davies1975,Unruh1976,Hawking1975,Crispino2008,BirrellDavies}.
In the right Rindler wedge, null coordinates satisfy
\(U=-\rho e^{-a\eta}\), \(V=\rho e^{a\eta}\), with \(\rho>0\):
boost time dilates the null variables. Fourier modes \(e^{-i\omega U}\)
are related to homogeneous half-line modes by the Mellin integral
\eqref{eq:sec6-v2-Mellin-integrals}, with the appropriate branch prescription.
For Schwarzschild, \(\rho\simeq\sqrt{8M(r-2M)}\) gives
\(ds_{(2)}^2\simeq-\kappa^2\rho^2dt^2+d\rho^2\), \(\kappa=1/(4M)\).
This is the near-horizon boost geometry, not the barrier maximum of
\eqref{eq:sec10-RW-v2}. The field-theoretic normalization and vacuum choice
are essential to the thermal particle interpretation.
Betzios, Gaddam and Papadoulaki use IHO partial-wave models in their
reformulation of 't~Hooft's black-hole S-matrix~\cite{BetziosGaddamPapadoulaki2016}.

A complementary connection is developed by Rajeev, Chakraborty and
Padmanabhan~\cite{RajeevChakrabortyPadmanabhan2018}. For a normal oscillator
with \(m(t)>0\), the change \(y=\sqrt{m(t)}\,q\), together with a total
derivative in the action, gives unit mass and
\begin{equation}
 \omega_{\rm eff}^2(t)=\omega^2(t)
 -\frac{1}{\sqrt{m(t)}}\frac{d^2\sqrt{m(t)}}{dt^2}.
\label{eq:sec10-time-dependent-inversion}
\end{equation}
This can be negative even when \(\omega^2(t)>0\). Their more general
transformations also reparametrize time, making the comparison with the
lens maps of Section~\ref{sec:hyperbolic-CN-TDSE} natural. At fixed time,
a time-dependent unitary map \(\psi'=U(t)\psi\) changes the Hamiltonian as
\(H'=UHU^{-1}+i\hbar\dot U U^{-1}\); this is not a stationary unitary
equivalence of the HO and IHO spectra. They also retain the coupled
background degree of freedom \(C(t)\) and derive a semiclassical
back-reaction prescription from total-energy conservation. That coupled
problem supplies physical information beyond an isolated quadratic mode.

Inflationary perturbations provide another time-dependent quadratic
example. As reviewed in Section~2.4 of Ref.~\cite{KumarMarto2026},
canonically rescaled scalar and tensor modes can acquire negative
effective squared frequency on sufficiently long wavelengths in a
quasi-de Sitter background. Their amplification and squeezing connect
hyperbolic dynamics with primordial spectra probed by CMB observations.
This is a mode-evolution problem with a time-dependent frequency;
predictions require the background evolution, initial state and
subsequent transfer of perturbations, rather than the stationary IHO
transmission coefficient alone.
The direct-sum inflation proposal of
Ref.~\cite{GaztanagaKumarMarto2026}, Sec.~7, further relates
parity-conjugate spacetime sectors to possible large-scale CMB
parity asymmetries; these predictions depend on its additional
state-space prescription.

Kumar and Marto's review \emph{Hawking radiation with pure
states}~\cite{KumarMarto2024Hawking} proposes a different state-space
construction: interior and exterior components are assigned to geometric
superselection sectors using discrete spacetime symmetries. They recover
a thermal number factor while interpreting the radiation in terms of pure
sector states. This interpretation depends on their direct-sum prescription;
it does not follow from the Mellin connection coefficients alone. Related
Einstein--Rosen and IHO/dual-IHO constructions appear in
Refs.~\cite{GaztanagaKumarMarto2026,KumarMarto2026}, the latter applying
them to a proposed treatment of the massive tensor ghost in quadratic
gravity. These proposals share hyperbolic mode structures with the present
analysis but impose additional quantization and gravitational assumptions;
the distinction from standard whole-line IHO scattering is discussed in
Section~\ref{sec:phase-space-wigner}.

\subsection{Dilation spectra and related symmetries}
\label{subsec:sec10-BK-v2}

The IHO--dilation rotation also links the scattering problem to
Berry--Keating \(xp\) dynamics~\cite{BerryKeating1999,BerryKeating99+}.
This established relation is used in the recent horizon and quadratic-gravity
papers~\cite{GaztanagaKumarMarto2026,KumarMarto2026}; the present triangle
places it alongside the FP threshold module, oscillator realization, and
their generating functions. The bare IHO/dilation operator has continuous
spectrum. Regularized \(xp\) models reproduce the smooth Riemann
zero-counting asymptotics, while identifying individual zeros requires
additional spectral conditions. Appendix~\ref{app:BK:IHO} distinguishes
these claims and reviews the indirect route to the zero-energy
inverse-square equation~\cite{SundaramBurgessODell2024,Camblong2000,BraatenPhillips2004,BawinCoon2003,EndresSteiner2010}.
Squaring the dilation operator and rescaling its eigenfunctions do not
establish a direct unitary equivalence of the IHO and inverse-square
Hamiltonians; their domains and scalar products must be specified.

The HO and IHO also realize the two Newton--Hooke signs~\cite{BacryLevyLeblond1968,DeromeDubois1972,Dubois1973,DuvalHorvathy2009}:
\(L=m(\dot x^2+\varepsilon x^2/R^2)/2\), \(\varepsilon=\pm1\), gives
\(\ddot x-\varepsilon x/R^2=0\). The negative sign is oscillatory and
the positive sign hyperbolic. Their translation and boost integrals contain
trigonometric or hyperbolic time dependence, whereas the free particle has
the time-independent translation integral \([H_0,p]=0\). This helps explain
the organizing role of its Fourier characters and threshold monomials.

\section{Discussion and outlook}
\label{sec:discussion-outlook}

The construction developed here extends the FP--HO conformal-bridge
program to the hyperbolic sector at the level of states and connection
data. Its contribution is the common derivation from the FP threshold
module and Fourier characters, with both resonant orientations and both
parity sectors retained. The canonical IHO--dilation rotation, oscillator
coherent states, IHO transmission coefficients and Cayley--Niederer
transformations are established ingredients; their coordinated role in
this dictionary is the content of the triangle in
Fig.~\ref{fig:FP-HO-IHO-triangle}.

\begin{figure}[!ht]
\centering
\resizebox{0.85\textwidth}{!}{%
\begin{tikzpicture}[
  >=Latex,
  thick,
  font=\small,
  boxHO/.style={
    draw=green!45!black,
    fill=green!7,
    rounded corners=3pt,
    align=center,
    inner sep=4pt,
    text width=3.3cm,
    minimum height=1.9cm
  },
  boxIHO/.style={
    draw=red!55!black,
    fill=red!5,
    rounded corners=3pt,
    align=center,
    inner sep=4pt,
    text width=3.3cm,
    minimum height=1.9cm
  },
  fpbox/.style={
    draw=blue!65!black,
    fill=blue!5,
    rounded corners=5pt,
    align=center,
    inner sep=4pt,
    text width=3.45cm,
    minimum height=1.95cm,
    line width=1.2pt
  },
  arrowlabel/.style={
    align=center,
    font=\scriptsize,
    fill=white,
    inner sep=2pt
  }
]

\node[fpbox] (FP) at (0,2.8)
{\(\mathrm{FP}\)\\[1mm]
\(H_0=\frac12 p^2,\ [H_0,p]=0\)\\
parabolic\\[1mm]
\(\{q^n,\ e^{ikq}\}\)};

\node[boxHO] (HO) at (-4.2,-1.4)
{\(\mathrm{HO}\)\\[1mm]
\(H_+=\frac12(p^2+q^2)\)\\
elliptic\\[1mm]
Fock and Glauber data};

\node[boxIHO] (IHO) at (4.2,-1.4)
{\(\mathrm{IHO}\)\\[1mm]
\(H_-=\frac12(p^2-q^2)\)\\
hyperbolic\\[1mm]
Gamow and scattering data};

\draw[->] (FP.south west) -- node[pos=0.50, left=7pt, arrowlabel]
{\(\mathrm{CBT}\) / Cayley\\
\(2iD\mapsto H_+\)\\
nonunitary in \(L^2\)\\
unitary to \(\mathcal F\)
\\generated by \(H_-\)}
(HO.north east);

\draw[->] (FP.south east) -- node[pos=0.53, right=9pt, arrowlabel]
{real metaplectic\\
\(R_{\pm\pi/4}\)\\
\(2D\mapsto\mp H_-\)
\\generated by \(H_+\)}
(IHO.north west);

\draw[->] (HO.east) -- node[midway, below=5pt, arrowlabel]
{analytic continuation\\
\(\omega\to\pm i\Omega\)\\
Weber connection}
(IHO.west);

\end{tikzpicture}%
}
\caption{The stationary FP--HO--IHO triangle in unit-frequency notation.
Here, \(\mathcal F\) denotes the Fock--Bargmann space.
The arrows organize transformations of the conformal generators and their
state realizations; they do not assert unitary equivalence of the three
Hamiltonians. The FP Fourier characters yield IHO generating states whose
Mellin decomposition gives the scattering connection data.}
\label{fig:FP-HO-IHO-triangle}
\end{figure}

The stationary result has two complementary parts. First, the rotations
in \eqref{eq:D-to-HIHO-quarter-rotations-sec4-v2} send the FP Jordan
states to the decaying and growing Gamow families, while the complex
FP--HO bridge sends them to the Fock tower. This makes explicit the
crossed use of the elliptic and hyperbolic generators and supplements
the spectral continuation $\omega\to\pm i\Omega$ with a direct state
map. Second, the transformed Fourier character supplies a generating
function with two expansions: Taylor coefficients give the Gamow
vectors, while Mellin coefficients give real-energy scattering data.
The common Gamma-function structure connects these two readings.
Neither the resonance energies nor the transmission probabilities are
claimed as new spectral formulas. The present construction connects
these data to the same FP threshold and Fourier-generating data used
in the HO bridge.

The $\mathfrak{osp}(1|2)$ description identifies the linear operators
that link the two parity sectors, beyond the quadratic algebra that
preserves each sector separately. Section~\ref{sec:coherent-perelomov-bogoliubov}
and Appendix~\ref{app:perelomov-packets} distinguish generalized
light-cone eigenstates from normalizable Glauber and Perelomov states.
At the level of evolution, Section~\ref{sec:hyperbolic-CN-TDSE} develops
the hyperbolic specialization of the projective-time construction of
Ref.~\cite{AlcalaPlyushchay2602.06378}, including its half-density,
chirp and propagator. A time-dependent intertwiner of Schr\"odinger
equations has a different meaning from a stationary spectral equivalence;
this distinction also governs comparison with time-dependent oscillator
inversion~\cite{RajeevChakrabortyPadmanabhan2018}.

The scattering interpretation is fixed by the incoming and outgoing
fluxes. The Fourier--Mellin connection matrix in
Section~\ref{sec:scattering-Fourier-Mellin-revised-v2} reproduces the
Weber asymptotics in Appendix~\ref{app:weber-iho}, including their
phase convention. Since the potential is unbounded below, these are
oscillatory asymptotic branches with varying local momentum, rather
than plane waves of constant momentum. Nonzero sub-barrier transmission
is compatible with exact Liouville transport: a sharp-energy Wigner
distribution obeys a star-eigenvalue equation and need not be supported
on the corresponding classical energy curve~\cite{BalazsVoros1990}.
This explains the coexistence of quantum transmission and classical
separatrices without assigning a crossing trajectory to a particle in
one classical sector. The sign projectors in the two conjugate
polarizations also do not supply four simultaneous quantum channel
labels.

Section~\ref{sec:physical-applications-final-style-v2}  places the local hyperbolic structure in a broader physical context and clarifies the role of global boundary conditions, field-mode normalization and the choice of quantum state space.
 In particular, the direct-sum horizon and quadratic-gravity
proposals~\cite{GaztanagaKumarMarto2026,KumarMarto2026,KumarMarto2024Hawking}
use assumptions beyond the whole-line IHO connection problem.
Their gravitational conclusions do not follow from the triangle alone.

A structural analogy can be drawn with holography. In the
AdS/CFT correspondence, bulk gravitational physics is related to a
boundary conformal field theory%
~\cite{Maldacena1998,GubserKlebanovPolyakov1998,Witten1998}.
For asymptotically AdS$_3$ gravity, the Brown--Henneaux boundary
conditions yield two Virasoro algebras of asymptotic symmetries, with
the AdS$_3$ isometry algebra as their global
subalgebra~\cite{BrownHenneaux1986}.
The FP--HO--IHO triangle is, of course, not a holographic duality.
Nevertheless, the free-particle realization plays a boundary-like
organizing role: it supplies simple parabolic data, while the
bridges provide the dictionary to elliptic and hyperbolic
realizations.

The linear-potential extension, $H_{\rm lin}=p^2/2+Fq$, has subsequently
been developed in Ref.~\cite{AlcalaPlyushchayAffine2026}. That work
constructs the regular accelerated-frame FP bridge and the singular
displaced-HO and displaced-IHO limits, and relates their spectral states
through the Airy transform. It places the linear-potential system in the
affine Schr\"odinger algebra, with its Heisenberg generators, beyond the
homogeneous quadratic conformal sector considered here.

\paragraph{Outlook.}

The boundary prescriptions suggest studying how time reversal or $PT$
exchanges resonant and anti-resonant sectors, particularly in
Berry--Keating-type spectral problems~\cite{BenderBrodyMuller2017,BenderBrody2018}.
The inverse-square reduction in Appendix~\ref{app:BK:IHO} also suggests
SUSY-type factorizations with a complex inverse-coordinate
superpotential. Their first-order factors are not ordinary adjoints
in the standard half-line scalar product; domains and any alternative
metric would have to be specified.

Finally, the trigonometric and hyperbolic scale factors in the
Cayley--Niederer maps suggest comparison with the logarithmic
derivatives of Darboux seeds for reflectionless and
P\"oschl--Teller systems. A precise relation to Darboux--KdV and
projective/Riccati geometry remains to be established.

\paragraph{Acknowledgments.}
The work was partially supported by the FONDECYT Project 1242046.


\appendix

\section{Order-eight property of the Cayley and \texorpdfstring{$\pi/4$}{pi/4}-rotation bridges}
\label{app:order-eight}

The finite-order properties of the stationary bridges require distinguishing
canonical matrices, metaplectic operators, and scalar phase conventions.
In the unit-frequency notation of Section~\ref{sec:conformal-bridges-revised-improved},
the complex canonical matrix underlying the FP--HO bridge is~\cite{AlcalaPlyushchay2602.06378}
\[
C=\frac{1}{\sqrt2}\begin{pmatrix}1&-i\\-i&1\end{pmatrix}
 =e^{-i\pi\sigma_1/4},\qquad C^4=-\mathbf1,\quad C^8=\mathbf1.
\]
It is induced by the analytically continued \(H_-\)-flow at \(t=i\pi/4\),
with \(U_C\propto e^{\pi H_-/(4\hbar)}\), as in
\eqref{eq:Cayley-operator-Hminus-sec4-v2}. The finite-order statement concerns
the formal complex canonical transformation and its action on the
Heisenberg generators. For the spectral exponential of self-adjoint
\(H_-\), \(U_C=c\,e^{\pi H_-/(4\hbar)}\) has no finite order:
\(U_C^8=c^8e^{2\pi H_-/\hbar}\ne\mathbf1\), since
\(\sigma(H_-)=\mathbb R\). Thus \(C^8=\mathbf1\) cannot be
iterated as a global Hilbert-space operator identity.

For the FP--IHO bridge, distinguish the real canonical rotation \(r_\theta\)
from its metaplectic lift \(R_\theta=e^{-i\theta H_+/\hbar}\), where
\(H_+=(p^2+q^2)/2\). The adjoint convention used in the main text gives
\[
r_\theta=\begin{pmatrix}\cos\theta&-\sin\theta\\
                         \sin\theta&\cos\theta\end{pmatrix},
\qquad r_{\pm\pi/4}^{\,4}=-\mathbf1,\quad r_{\pm\pi/4}^{\,8}=\mathbf1.
\]
Since \(H_+|n\rangle=\hbar(n+1/2)|n\rangle\), the lift instead satisfies
\[
R_{\pm\pi/4}^{\,8}|n\rangle=e^{\mp2\pi i(n+1/2)}|n\rangle=-|n\rangle,
\qquad R_{\pm\pi/4}^{\,16}=\mathbf1.
\]
This central sign is invisible in the adjoint action. Removing the oscillator
zero-point phase gives the normalized fractional Fourier convention
\[
\mathcal F_\theta=e^{i\theta/2}R_\theta=e^{-i\theta a^+a^-},
\qquad \mathcal F_{\pm\pi/4}^{\,8}=\mathbf1.
\]
Thus both canonical bridges have order eight; the real rotation has an
order-sixteen metaplectic lift, while its phase-normalized representative
has order eight.

\section{Weber functions, scattering currents, and coherent-state overlaps}
\label{app:weber-iho}

This appendix makes explicit the relation between the Fourier--Mellin
coefficients of Section~\ref{sec:scattering-Fourier-Mellin-revised-v2} and
coordinate-space scattering currents. The Weber and Wigner calculations are
standard IHO constructions~\cite{Barton1990,BalazsVoros1990,DLMF12p2}; we give
them in a common convention to specify the boundary conditions and the meaning
of transmission. We then distinguish the scattering matrix from the
state-dependent overlaps with HO coherent states.

\subsection{From an incoming Mellin mode to coordinate-space currents}
\label{subsec:appB-flux}

For the unit-mass Hamiltonian \(H=p^2/2-\Omega^2q^2/2\), introduce
\begin{equation}
x=\sqrt{\frac{\Omega}{\hbar}}q,\quad
k=\frac{p}{\sqrt{\hbar\Omega}},\quad
\xi=\frac{k+x}{\sqrt2},\quad \eta=\frac{k-x}{\sqrt2},\quad
\epsilon=\frac{E}{\hbar\Omega},\quad s=\frac12-i\epsilon.
\label{eq:appB-variables}
\end{equation}
Then \([x,k]=[\xi,\eta]=i\), and the Schr\"odinger equation is
\(\frac12(-\partial_x^2-x^2)\psi=\epsilon\psi\).
The left-incident Mellin mode is \(f_\epsilon(\eta)=\Theta(\eta)\eta^{s-1}\).
Here we omit the common energy-normalization factor: the resulting asymptotic
incident current will be one in dimensionless time \(\theta=\Omega t\).

The unitary transformation from \(\eta\) to \(x\) can be chosen with kernel
\begin{equation}
K(x,\eta)=\frac{2^{1/4}}{\sqrt{2\pi}}
\exp\!\left[i\left(\frac{x^2}{2}+\sqrt2 x\eta+\frac{\eta^2}{2}\right)\right].
\label{eq:appB-kernel}
\end{equation}
Indeed, \((-i\partial_x-x)K/\sqrt2=\eta K\), and
\(\int_{\mathbb R}K(x,\eta)K(x',\eta)^*d\eta=\delta(x-x')\).
Transforming \(f_\epsilon\) gives the exact Weber solution
\begin{equation}
\begin{split}
\psi_{L,\epsilon}(x)
&=\frac{2^{1/4}}{\sqrt{2\pi}}e^{ix^2/2}
\int_0^\infty \eta^{s-1}e^{i(\eta^2/2+\sqrt2 x\eta)}d\eta\\
&=\frac{2^{1/4}\Gamma(s)e^{i\pi s/4}}{\sqrt{2\pi}}
D_{-s}(\sqrt2 e^{-i\pi/4}x).
\end{split}
\label{eq:appB-left-Weber}
\end{equation}
The oscillatory integral is evaluated as a boundary value from a damped
integral. For example, inserting \(e^{-\delta\eta^2}\) and taking
\(\delta\to0^+\) fixes the continuation in the second line. Its Weber index
\(-s=-1/2+i\epsilon\) agrees with \eqref{eq:sec6-v2-Weber}.

For fixed \(\epsilon\), put \(F(X)=X^2/2+\epsilon\log X\), \(X>0\).
At \(x=-X\to-\infty\), the integration phase has a saddle at
\(\eta=\sqrt2X\) and an endpoint at \(\eta=0\). They supply the incident
and reflected waves, respectively. At \(x\to+\infty\) only the endpoint
contributes, giving a transmitted wave and no incoming wave from the right.
The leading asymptotics are
\begin{align}
\psi_{L,\epsilon}(-X)&\sim C_\epsilon X^{-1/2}
\left[e^{-iF(X)}+r_{\rm coord}(E)e^{iF(X)}\right],\notag\\
\psi_{L,\epsilon}(x)&\sim C_\epsilon t_{\rm coord}(E)x^{-1/2}e^{iF(x)},
\qquad C_\epsilon=e^{i\pi/4}2^{-i\epsilon/2}.
\label{eq:appB-asymptotics}
\end{align}
Here \(t_{\rm coord}(E)\) and \(r_{\rm coord}(E)\) are the transmitted and
reflected amplitudes relative to the incident wave in the coordinate-space
asymptotics \eqref{eq:appB-asymptotics}. The label ``coord'' denotes this
choice of asymptotic basis; these coefficients depend on energy, not on
position. By contrast, \(t(E)\) and \(r(E)\) are the Fourier--Mellin
connection amplitudes of \eqref{eq:sec6-v2-tr-amplitudes}. With the
dimensionless Mellin convention \(\varphi_0=0\), their relation is
\begin{equation}
\begin{aligned}
t_{\rm coord}(E)&=e^{i(\epsilon\log2-\pi/4)}t(E),&
r_{\rm coord}(E)&=e^{i(\epsilon\log2-\pi/4)}r(E),\\
t(E)&=\frac{\Gamma(s)}{\sqrt{2\pi}}e^{i\pi s/2},&
r(E)&=\frac{\Gamma(s)}{\sqrt{2\pi}}e^{-i\pi s/2}.
\end{aligned}
\label{eq:appB-channel-phase}
\end{equation}
They describe the same scattering process in two channel-phase conventions:
the displayed common phase comes from the asymptotic phases chosen in
\eqref{eq:appB-asymptotics}. Thus \(|t_{\rm coord}|^2=|t|^2\),
\(|r_{\rm coord}|^2=|r|^2\), and \(r_{\rm coord}/t_{\rm coord}=r/t\).
The \(X^{-1/2}\) amplitudes and quadratic/logarithmic phases in
\eqref{eq:appB-asymptotics} are the large-distance Weber/WKB channels,
not plane waves.

For dimensionless time \(\theta\), the current is
\(j_x=\operatorname{Im}(\psi^*\partial_x\psi)\).
The minus sign in \(\partial_x=-\partial_X\) on the left identifies
\(e^{-iF(X)}\) as incoming and \(e^{iF(X)}\) as reflected. Since
\(|C_\epsilon|=1\), their asymptotic currents and that on the right are
\begin{equation}
j_{\rm in}=1,\qquad j_{\rm ref}=-|r_{\rm coord}|^2,
\qquad j_{\rm tr}=|t_{\rm coord}|^2,
\qquad
\frac{j_{\rm tr}}{j_{\rm in}}=T(E),\quad
\frac{|j_{\rm ref}|}{j_{\rm in}}=R(E).
\label{eq:appB-current-ratios}
\end{equation}
Restoring the coordinate scale and energy normalization multiplies all three
currents by the same positive factor. For \(\epsilon<0\), the interval
\(|x|<\sqrt{-2\epsilon}\) is classically forbidden, yet the same fixed-energy
solution carries nonzero transmitted current to its right. This is the
coordinate-space tunnelling interpretation used in the main text.

For a normalized, purely left-incident packet in the energy-normalized basis,
with spectral amplitude \(a(E)\), the corresponding probabilities are
\begin{equation}
P_{\rm tr}=\int_{\mathbb R}|a(E)|^2T(E)\,dE,\qquad
P_{\rm ref}=\int_{\mathbb R}|a(E)|^2R(E)\,dE,
\qquad \int_{\mathbb R}|a(E)|^2dE=1.
\label{eq:appB-packet-probabilities}
\end{equation}
The fixed-energy effect in \eqref{eq:appB-current-ratios} therefore also has
the usual asymptotic packet interpretation.

\subsection{Wigner support and the same current ratios}
\label{subsec:appB-wigner}

Let \(h=(k^2-x^2)/2=\xi\eta\) denote the dimensionless classical symbol,
distinct from the quantum eigenvalue \(\epsilon\). Since the momentum
conjugate to \(\eta\) is \(-\xi\), the Wigner transform of the same
incoming mode can be written as
\begin{equation}
\begin{split}
\mathcal W_\epsilon(\xi,\eta)
&=\frac{1}{2\pi}\int_{\mathbb R}e^{i\xi y}
f_\epsilon(\eta+y/2)f_\epsilon(\eta-y/2)^*\,dy
=\Theta(\eta)w_\epsilon(h),\\
w_\epsilon(h)
&=\frac1\pi\int_{-1}^1
\frac{\exp\!\left[2ihz-i\epsilon\log\frac{1+z}{1-z}\right]}
{\sqrt{1-z^2}}\,dz,\qquad
\int_{\mathbb R}w_\epsilon(h)\,dh=1.
\end{split}
\label{eq:appB-Wigner-integral}
\end{equation}
For \(\eta>0\), this follows by setting \(y=2\eta z\); for \(\eta<0\)
the two factors have no common support. The last identity fixes the incident
flux; it is not a normalization of the stationary state over the phase plane.
The integrals over \(h\) carry their oscillatory boundary-value prescription.
The distribution is not confined to \(h=\epsilon\), nor to negative \(h\)
when \(\epsilon<0\). This is the support mechanism identified by
Balazs and Voros~\cite{BalazsVoros1990}.

Metaplectic covariance allows the same distribution to be expressed in
\((x,k)\). Its support condition is \(k>x\). On the right, \(x>0\),
the current is therefore
\begin{equation}
j_{\rm tr}=\int_x^\infty k\,
w_\epsilon\!\left(\frac{k^2-x^2}{2}\right)dk
=\int_0^\infty w_\epsilon(h)\,dh.
\label{eq:appB-Wigner-current}
\end{equation}
On the left, fix \(x=-X<0\). The support condition is now \(k>-X\).
Splitting the current into positive- and negative-momentum contributions,
and changing variable at this fixed position to
\(h=(k^2-X^2)/2\), with \(dh=k\,dk\), gives
\[
\begin{aligned}
j_+(-X)&=\int_0^\infty k\,w_\epsilon\!\left(\frac{k^2-X^2}{2}\right)dk
       =\int_{-X^2/2}^{\infty}w_\epsilon(h)\,dh,\\
j_-(-X)&=\int_{-X}^{0} k\,w_\epsilon\!\left(\frac{k^2-X^2}{2}\right)dk
       =-\int_{-X^2/2}^{0}w_\epsilon(h)\,dh.
\end{aligned}
\]
The minus sign in the second line comes from the reversed limits:
as \(k\) increases from \(-X\) to zero, \(h\) decreases from zero to
\(-X^2/2\). Their far-left limits identify the incident and reflected currents:
\[
\begin{aligned}
j_{\rm in}&=\lim_{X\to\infty}j_+(-X)
          =\int_{-\infty}^{\infty}w_\epsilon(h)\,dh=1,\\
j_{\rm ref}&=\lim_{X\to\infty}j_-(-X)
           =-\int_{-\infty}^{0}w_\epsilon(h)\,dh.
\end{aligned}
\]
Thus the integration intervals are ranges of the classical symbol \(h\),
not spatial intervals. The spatial asymptotic limit has already been taken
in these expressions. They are consistent with
\(j_+(-X)+j_-(-X)=\int_0^\infty w_\epsilon(h)\,dh=j_{\rm tr}\).
Using \(z=\tanh y\) in \eqref{eq:appB-Wigner-integral} gives
\begin{equation}
\int_0^\infty w_\epsilon(h)\,dh
=\frac12+\frac1\pi\int_0^\infty
\frac{\sin(2\epsilon y)}{\sinh y}\,dy
=\frac12[1+\tanh(\pi\epsilon)]=T(E),
\label{eq:appB-Wigner-T}
\end{equation}
and the negative-\(h\) integral gives \(R(E)\).
Thus exact transport along classical characteristics and nonzero
fixed-energy transmission coexist. The support at positive classical
\(h\) is a property of the quantum Wigner distribution, not a conversion
of its energy eigenvalue from negative to positive.

\subsection{Extracting channel amplitudes from coherent-state decompositions}
\label{subsec:appB-coherent}

Choose energy-normalized incoming and outgoing IHO bases with the phases
and sign ordering of Section~\ref{sec:scattering-Fourier-Mellin-revised-v2}.
Their connection is
\begin{equation}
|E,\sigma;\mathrm{in}\rangle
=\sum_{\tau=\pm}|E,\tau;\mathrm{out}\rangle S_{\tau\sigma}(E),
\qquad
\langle E,\tau;\mathrm{out}|E',\sigma;\mathrm{in}\rangle
=\delta(E-E')S_{\tau\sigma}(E).
\label{eq:appB-basis-connection}
\end{equation}
For an HO coherent state define
\(C^{a}_\sigma(E;\alpha)=\langle E,\sigma;a|\alpha\rangle\),
\(a=\mathrm{in},\mathrm{out}\). Expanding the same state in either basis gives
\begin{equation}
\begin{pmatrix}C_+^{\rm out}(E;\alpha)\\C_-^{\rm out}(E;\alpha)\end{pmatrix}
=\begin{pmatrix}t(E)&r(E)\\r(E)&t(E)\end{pmatrix}
\begin{pmatrix}C_+^{\rm in}(E;\alpha)\\C_-^{\rm in}(E;\alpha)\end{pmatrix}.
\label{eq:appB-coherent-connection}
\end{equation}
These coefficients can be evaluated as the coordinate-space overlaps
\eqref{eq:sec6-v2-coherent-Weber-overlap}. Equivalently, if
\(g_\alpha(v)=\langle v|\alpha\rangle\) in the incoming light-cone
representation, then
\begin{equation}
C^{\rm in}_\sigma(E;\alpha)
=N_E\int_0^\infty v^{-1/2+i\epsilon}g_\alpha(\sigma v)\,dv,
\qquad \sigma=\pm,
\label{eq:appB-coherent-Mellin}
\end{equation}
with the same fixed reference scale in the powers as in Section~6.
For a normalizable HO coherent state, \(g_\alpha\) is a Gaussian quadrature
wavefunction. Both incoming half-lines generally contribute. The two outgoing
coefficients in \eqref{eq:appB-coherent-connection} are consequently not,
in general, proportional to \((t,r)\).

To isolate the column describing incidence from the left, retain the
incoming \(+\) component:
\begin{equation}
|\alpha;+\mathrm{in}\rangle
:=\Theta(\hat u_-)|\alpha\rangle
=\int_{\mathbb R}C_+^{\rm in}(E;\alpha)
|E,+;\mathrm{in}\rangle\,dE.
\label{eq:appB-coherent-projection}
\end{equation}
This is a projected component, generally no longer an HO coherent state;
its normalization is immaterial to the following ratios. Its outgoing
spectral weights satisfy
\begin{equation}
\widetilde C_+^{\rm out}(E;\alpha)
=t(E)C_+^{\rm in}(E;\alpha),\qquad
\widetilde C_-^{\rm out}(E;\alpha)
=r(E)C_+^{\rm in}(E;\alpha).
\label{eq:appB-coherent-extraction}
\end{equation}
Dividing out the common incoming weight wherever nonzero extracts the two
amplitudes. The opposite projection gives the other column of \(S\).
Thus the Weber/Mellin connection supplies the universal scattering matrix,
while the coherent state supplies the spectral envelope.

\section{Dilation, Berry--Keating structures, and the inverse-square-potential correspondence}
\label{app:BK:IHO}

The IHO--dilation relation places Mellin scattering, Berry--Keating
spectral questions, and the inverse-square potential (ISP) in a common
hyperbolic setting~\cite{BerryKeating1999,BerryKeating99+,SundaramBurgessODell2024,Camblong2000,BraatenPhillips2004,BawinCoon2003,EndresSteiner2010,SierraRodriguezLaguna2011,Sierra2019}.
The real rotation through \(\pi/4\) is an established canonical relation.
Gazta\~naga, Kumar and Marto discuss this rotation and Berry--Keating
conditions in Sec.~3 of Ref.~\cite{GaztanagaKumarMarto2026}, with their
direct-sum construction in Sec.~5.3; the IHO/dual-IHO treatment of
Ref.~\cite{KumarMarto2026} develops related spectral conditions in
Secs.~2.2 and~3.1. Those constructions introduce boundary or sector
prescriptions beyond the unregularized whole-line scattering problem.
Here we distinguish its continuous spectrum, the regularized smooth
zero-counting law, and proposals to select individual Riemann zeros. The
last do not follow from the canonical rotation alone. The further ISP
correspondence passes through a squared equation and a change of scalar
product, rather than a direct identity of the original Hamiltonians.

\noindent
The notation in this Appendix follows the conventional Berry--Keating/ISP usage. Thus the
IHO coordinate is denoted by \(x\), and the \(\pi/4\)-rotated dilation variables by \(X,P\). 
They are the same rotated canonical variables that, in the main text, are described in terms
of the light-cone variables \(u_\pm\).

\subsection{IHO and the dilation/Berry--Keating operator}

In units with mass and oscillator frequency equal to one, let
\(H_-=(p^2-x^2)/2\), with \([x,p]=i\hbar\).
Introduce the real canonical variables
\(X=(p+x)/\sqrt{2}\) and \(P=(p-x)/\sqrt{2}\),
which satisfy \([X,P]=i\hbar\).
Classically, this gives \(H_{-}=XP\).  Quantum mechanically the metaplectic lift of the same \(\pi/4\) rotation gives the symmetrized operator
\begin{equation}
H_{-}\quad \longleftrightarrow \quad
H_{\rm BK}:=\frac{1}{2}(XP+PX).
\label{app:BK:HBK}
\end{equation}
Thus \(H_{\rm BK}=2D\) in the rotated canonical pair \((X,P)\): the IHO and dilation operators represent the same hyperbolic metaplectic generator in different polarizations.

In the \(X\)-representation,
\(H_{\rm BK}=-i\hbar(X\,d/dX+1/2)\),
so its generalized eigenfunctions are Mellin powers.
This is the same multiplicative-character structure used
in the light-cone treatment of IHO scattering.

\subsection{Berry--Keating regularization and the smooth counting law}

The importance of the Berry--Keating proposal is not that it supplies a complete Hilbert--Polya operator.  Its role is more modest and more robust: a regularized classical \(xp\) Hamiltonian reproduces the smooth \(E\log E\) part of the Riemann zero-counting function.  This is precisely the part one expects from scale-invariant hyperbolic motion.

For the classical Hamiltonian \(H=xp\), the positive-energy trajectory in the first quadrant is \(p=E/x\).  With the cutoffs \(x\geq \ell_x\), \(p\geq \ell_p\), and \(\ell_x\ell_p=2\pi\hbar\), the truncated phase-space area for \(E\geq 2\pi\hbar\) is
\begin{equation}
A(E)=\int_{\ell_x}^{E/\ell_p}\left(\frac{E}{x}-\ell_p\right)dx
=E\log\left(\frac{E}{2\pi\hbar}\right)-E+2\pi\hbar .
\label{app:BK:area}
\end{equation}
Dividing by a Planck cell gives the raw count
\begin{equation}
N_{\rm raw}(E)=\frac{E}{2\pi\hbar}\left[\log\left(\frac{E}{2\pi\hbar}\right)-1\right]+1.
\label{app:BK:rawcount}
\end{equation}
The leading \(E\log E\) and \(-E\) terms match the smooth Riemann--von Mangoldt behavior after the usual identification of the energy scale.  The constant term is not fixed by the area alone; the familiar smooth counting formula contains the constant \(7/8\), so one needs the additional semiclassical phase shift \(\delta_{\rm M}=-1/8\).  This is often described as a Maslov-type correction.  Unlike the ordinary harmonic oscillator, however, the bare \(xp\) motion is open and hyperbolic, so this constant should be understood as part of a chosen regularized semiclassical closure, not as a canonical turning-point index of the unregularized flow.

The logarithm in \eqref{app:BK:area} has a simple origin: it is the integral \(\int dx/x\) produced by the scale-invariant core of \(xp\).  The regularization breaks exact continuous scale invariance, but the leading logarithmic trace of that symmetry remains.

\subsection{The indirect IHO--BK--ISP chain}

The relation between the IHO and the inverse-square potential is real but indirect.  Starting from \eqref{app:BK:HBK}, the Berry--Keating eigenvalue equation \(H_{\rm BK}\phi=\widehat E\phi\) is first order.  In this subsection the positive dilation coordinate is denoted by \(Q>0\), to match the standard inverse-square notation. With \(\hbar=1\), it reads
\begin{equation}
-i\left(Q\frac{d}{dQ}+\frac{1}{2}\right)\phi(Q)=\widehat E\phi(Q),
\qquad
\phi(Q)\propto Q^{-1/2+i\widehat E}
\label{app:BK:firstorder}
\end{equation}
up to the sign convention for \(\widehat E\).  To obtain a second-order Schr\"odinger-type equation one squares the BK operator.  At the differential-equation level this gives an Euler equation, and the substitution \(\phi(Q)=\chi(Q)/Q\) removes the first-derivative term.  One obtains
\begin{equation}
-\chi''(Q)-\frac{\widehat E^{2}+1/4}{Q^{2}}\chi(Q)=0.
\label{app:BK:ISPzero}
\end{equation}
This is the zero-energy Schr\"odinger equation for an attractive inverse-square potential.  For \(H_{\rm ISP}=-d^{2}/dQ^{2}-g/Q^{2}\), the coupling is \(g=\widehat E^{2}+1/4\); if the kinetic term is instead \(-\tfrac12d^{2}/dQ^{2}\), the convention is \(2g=\widehat E^{2}+1/4\).  For real \(\widehat E\ne0\) this is supercritical, while \(\widehat E=0\) gives the critical coupling.  Thus the IHO/BK energy label becomes the inverse-square coupling, with ISP energy fixed to zero. Squaring combines the \(\widehat E\) and \(-\widehat E\) branches, so recovering a specified first-order BK eigenvalue requires retaining its sign. At \(\widehat E=0\), the squared equation additionally admits the logarithmic solution \(\chi\propto Q^{1/2}\ln Q\).

The chain should therefore be read as
\begin{equation}
\mathrm{IHO}
\xrightarrow{\ \pi/4\ \mathrm{rotation}\ }
\mathrm{BK}
\xrightarrow{\ \mathrm{squaring}\ }
H_{\rm BK}^{2}
\xrightarrow{\ \phi=\chi/Q\ }
\mathrm{zero\mbox{-}energy\ ISP}.
\label{app:BK:chain}
\end{equation}

\subsection{Supercritical ISP, boundary data, and RG flow}

The attractive inverse-square Hamiltonian on the half-line,
\begin{equation}
H_{\rm ISP}=-\frac{d^{2}}{dr^{2}}-\frac{g}{r^{2}},
\qquad r>0,
\label{app:BK:HISP}
\end{equation}
is singular at the origin.  In the supercritical regime \(g>1/4\) one writes \(\nu=(g-1/4)^{1/2}\), and the two independent near-origin behaviors are
\begin{equation}
\psi(r)\sim \alpha r^{1/2-i\nu}+\beta r^{1/2+i\nu}
=r^{1/2}\left(\alpha e^{-i\nu\ln r}+\beta e^{i\nu\ln r}\right).
\label{app:BK:supercritical}
\end{equation}
Both branches have the same magnitude and oscillate in \(\ln r\).  The differential equation alone does not select a unique physical solution.  One must specify short-distance boundary data, equivalently a self-adjoint extension in the conservative problem.

A cutoff description makes the renormalization-group structure transparent.  At \(r=\epsilon\), introduce the logarithmic boundary parameter
\begin{equation}
\Lambda(\epsilon):=\epsilon\frac{\psi'(\epsilon)}{\psi(\epsilon)}-\frac{1}{2}.
\label{app:BK:Lambda}
\end{equation}
At zero ISP energy, keeping the ratio \(\alpha/\beta\) fixed gives the RG equation below. It also gives the leading short-distance flow at finite ISP energy. For self-adjoint boundary data, \(\Lambda\) is real and its solution is parametrized by a positive length \(\epsilon_*\):
\begin{equation}
\frac{d\Lambda}{d\ln\epsilon}=-(\nu^{2}+\Lambda^{2}),
\qquad
\Lambda(\epsilon)=\nu\cot\left[\nu\ln\left(\frac{\epsilon}{\epsilon_*}\right)\right].
\label{app:BK:RG}
\end{equation}
Thus the flow is periodic in \(
\ln\epsilon\).  The multiplicative period \(\epsilon\mapsto e^{\pi/\nu}\epsilon\) is the familiar limit cycle and expresses discrete scale invariance.  Equivalently, the renormalized theory contains an RG-invariant scale, often written as a length \(r_0\) or a momentum \(\kappa_*\).  The bound-state momenta then form a geometric tower \(\kappa_n=\kappa_* e^{-n\pi/\nu}\), with energies proportional to \(-\kappa_n^2\).

For a closed Hermitian problem the self-adjoint-extension parameter can be described by fixing the relative phase of the two near-origin branches, for example \(|\alpha|=|\beta|\) with \(\alpha/\beta=e^{-2i\theta}\).  The RG language and the self-adjoint-extension language are two ways of encoding the same short-distance ambiguity: the former emphasizes running boundary data, the latter emphasizes the admissible domains of the Hamiltonian.

\subsection{Operator-domain caveat}

The formal chain \eqref{app:BK:chain} must not be overinterpreted as a unitary equivalence of all self-adjoint operators involved.  There are two separate issues.

First, an operator and its square generally have different domains: \(D(A^{2})=\{\psi\in D(A)\,|\, A\psi\in D(A)\}\).  Thus passing from \(H_{\rm BK}\) to \(H_{\rm BK}^{2}\) is not only an algebraic operation on differential expressions; it also changes the domain question.

Second, the step \(\phi=\chi/Q\) in \eqref{app:BK:chain} is not unitary from \(L^{2}(0,\infty;dQ)\) to the standard Schr\"odinger space \(L^{2}(0,\infty;dQ)\).  Indeed, \(\|\phi\|^{2}=\int_{0}^{\infty}|\chi(Q)|^{2}dQ/Q^{2}\).  The natural image space for \(\chi\) is therefore weighted, not the usual half-line Hilbert space.  Consequently, the standard inverse-square Hamiltonian in \(L^{2}(0,\infty;dQ)\) has its own self-adjoint-extension problem, and its domains are not automatically inherited from the squared BK operator.

The IHO on the full line is already a well-defined self-adjoint scattering Hamiltonian; the ISP short-distance ambiguity is not a pathology transported back to the IHO operator itself. Rather, under the duality it appears as a choice of state-selection or boundary data.


\section{Perelomov states, squeezed packets, and spectral states}
\label{app:perelomov-packets}

With \(a^-=(Q+iP)/\sqrt{2\hbar}\), \(a^+=(a^-)^\dagger\), the quadratic
generators are
\begin{equation}
 K_0=\frac12\left(a^+a^-+\frac12\right),\qquad
 K_+=\frac12(a^+)^2,\qquad K_-=\frac12(a^-)^2.
\label{eq:sec9-K-generators-revised}
\end{equation}
They obey \([K_0,K_\pm]=\pm K_\pm\), \([K_-,K_+]=2K_0\), and preserve
parity, with \(H_{\rm IHO}=-\hbar\Omega(K_++K_-)\).
For \(k=1/4,3/4\), with lowest-weight vectors
\(|k,0\rangle=|0\rangle,|1\rangle\), respectively, the normalized
Perelomov states are~\cite{Perelomov}
\begin{equation}
\begin{aligned}
 |\zeta;k\rangle&=\mathcal D(\xi)|k,0\rangle
 =(1-|\zeta|^2)^k e^{\zeta K_+}|k,0\rangle,\\
 \mathcal D(\xi)&=e^{\xi K_+-\xi^*K_-},\qquad
 \zeta=e^{i\arg\xi}\tanh|\xi|,\quad |\zeta|<1.
\end{aligned}
\label{eq:sec9-Perelomov-normal-form-revised}
\end{equation}
IHO evolution is a squeeze with \(\xi=i\Omega t\):
\begin{equation}
 e^{-itH_{\rm IHO}/\hbar}=\mathcal D(i\Omega t),\qquad
 \zeta(t)=i\tanh(\Omega t).
\label{eq:sec9-IHO-SU11-evolution-revised}
\end{equation}
It sends \(|0\rangle\) to a centered Gaussian, and \(|1\rangle\) to an
odd, non-Gaussian squeezed state. The latter is a linear polynomial times
a Gaussian in coordinate space.

A general pure Gaussian additionally uses a Heisenberg--Weyl displacement:
\begin{equation}
 |\alpha,\zeta\rangle=D_{\rm HW}(\alpha)\mathcal D(\xi)|0\rangle,
 \qquad D_{\rm HW}(\alpha)=e^{\alpha a^+-\alpha^*a^-}.
\label{eq:sec9-displaced-squeezed-explicit-revised}
\end{equation}
The constructions discussed in Section~\ref{sec:coherent-perelomov-bogoliubov}
can be classified as follows:
\begin{center}
\small
\begin{tabular}{>{\raggedright\arraybackslash}p{0.34\textwidth}
                >{\raggedright\arraybackslash}p{0.60\textwidth}}
\toprule
\textbf{Object} & \textbf{Type and role}\\
\midrule
HO Glauber state & Normalizable \(a^-\)-eigenstate; normalized FP--HO image of a plane wave.\\
IHO light-cone eigenstate & Generalized \(u_\pm\)-eigenstate; Taylor and Mellin expansions probe Gamow and scattering data.\\
Even/odd Perelomov states & Normalizable squeezed \(|0\rangle\)/\(|1\rangle\); only the even family is Gaussian.\\
Displaced squeezed vacuum & Pure Gaussian packet with a classical hyperbolic center.\\
IHO scattering state & Delta-normalized real-energy eigenstate of \(H_{\rm IHO}\).\\
IHO Gamow state & Resonant generalized state at an analytically continued complex energy.\\
\bottomrule
\end{tabular}
\end{center}
The scattering states, including \(E=0\), are not Perelomov states.
For packet overlaps and the action of the scattering matrix, see
Appendix~\ref{subsec:appB-coherent}.

\end{document}